\documentclass[a4paper,11pt]{article}
\usepackage{jcappub}
\usepackage{amsmath}
\usepackage{bbm}

\usepackage{graphicx}

\makeatletter
\gdef\@fpheader{}
\makeatother

\newlength{\wsingfig}
\newlength{\wdblefig}
\DeclareMathOperator{\Tr}{Tr}
\DeclareMathOperator{\diag}{diag}
\DeclareMathOperator{\Ei}{Ei}

\newcommand{\Lambert}[1]{\mathrm{W}_{\negthinspace {#1}}}

\newcommand{\ud}{\mathrm{d}}

\newcommand{\N}[1]{N_{\mathrm {#1}}}

\newcommand{\us}{\mathrm{s}}
\newcommand{\uini}{\mathrm{ini}}
\newcommand{\uend}{\mathrm{end}}
\newcommand{\unuc}{\mathrm{nuc}}
\newcommand{\ureh}{\mathrm{reh}}
\newcommand{\cs}{c_{\us}}

\newcommand{\MeV}{\mathrm{MeV}}

\newcommand{\csa}{{c_{\us}}_{_0}}
\newcommand{\csb}{{c_{\us}}_{_1}}

\newcommand{\order}[1]{\mathcal{O}\!\left(#1\right)}

\newcommand{\ur}{\mathrm{r}}

\renewcommand{\d}{\mathrm{d}}

\newcommand{\Mm}{\hat{M}}
\newcommand{\Wm}{\hat{W}}
\newcommand{\Rm}{\hat{R}}
\newcommand{\gm}{\hat{g}}

\newcommand{\Tm}{\hat{T}}

\newcommand{\Id}{\mathbbm 1}

\newcommand{\calT}{\mathcal{T}}

\newcommand{\gt}{\tilde{g}}
\newcommand{\hT}{\tilde{h}}
\newcommand{\mut}{\tilde{\mu}}
\newcommand{\Ht}{\tilde{H}}

\newcommand{\at}{\tilde{a}}
\newcommand{\nt}{\tilde{n}}
\newcommand{\etat}{\tilde{\eta}}

\newcommand{\rhob}{\bar{\rho}}
\newcommand{\Pb}{\bar{P}}
\newcommand{\Hb}{\bar{H}}
\newcommand{\Hbend}{\bar{H}_{\uend}}

\newcommand{\rhobr}{\rhob_\ur}
\newcommand{\rhobini}{\rhob_\uini}
\newcommand{\rhobreh}{\rhob_\ureh}

\newcommand{\Gammab}{\bar{\Gamma}}

\newcommand{\Mlambda}{m_\lambda}
\newcommand{\Mp}{M_{\mathrm{Pl}}}
\newcommand{\mS}{\mathcal{S}}

\newcommand{\lsim}   {\mathrel{\mathop{\kern 0pt \rlap
  {\raise.2ex\hbox{$<$}}}
  \lower.9ex\hbox{\kern-.190em $\sim$}}}
\newcommand{\gsim}   {\mathrel{\mathop{\kern 0pt \rlap
  {\raise.2ex\hbox{$>$}}}
  \lower.9ex\hbox{\kern-.190em $\sim$}}}

\begin{document}


\title{Cascading dust inflation in Born-Infeld gravity}

\author[a,b]{Jose Beltr\'an Jim\'enez,}
\author[c]{Lavinia Heisenberg,}
\author[d,e]{Gonzalo Olmo,}
\author[b]{and Christophe Ringeval}

\affiliation[a]{CPT, Aix Marseille Universit\'e, UMR 7332, 13288
  Marseille,  France.}

\affiliation[b]{Centre for Cosmology, Particle Physics and Phenomenology,
  Institute of Mathematics and Physics, Louvain University, 2 Chemin
  du Cyclotron, 1348 Louvain-la-Neuve, Belgium.}

\affiliation[c]{Institute for Theoretical Studies, ETH Zurich, Clausiusstrasse 47, 8092 Zurich, Switzerland.}

\affiliation[d]{Depto. de F\'{i}sica Te\'{o}rica \& IFIC, Universidad
  de Valencia - CSIC,Calle Dr. Moliner 50, Burjassot 46100, Valencia,
  Spain.}

\affiliation[e]{Depto. de F\'isica, Universidade Federal da Para\'\i
  ba, Cidade Universit\'aria, s/n - Castelo Branco, 58051-900 Jo\~ao
  Pessoa, Para\'\i ba, Brazil.}

\emailAdd{jose.beltran@cpt.univ-mrs.fr} 
\emailAdd{lavinia.heisenberg@eth-its.ethz.ch}
\emailAdd{gonzalo.olmo@csic.es}
\emailAdd{christophe.ringeval@uclouvain.be}

\date{\today}

\abstract{In the framework of Born-Infeld inspired gravity theories,
  which deviates from General Relativity (GR) in the high curvature
  regime, we discuss the viability of Cosmic Inflation without scalar
  fields. For energy densities higher than the new mass scale of the
  theory, a gravitating dust component is shown to generically induce
  an accelerated expansion of the Universe. Within such a simple
  scenario, inflation gracefully exits when the GR regime is
  recovered, but the Universe would remain matter dominated. In order
  to implement a reheating era after inflation, we then consider
  inflation to be driven by a mixture of unstable dust species
  decaying into radiation. Because the speed of sound gravitates
  within the Born-Infeld model under consideration, our scenario ends
  up being predictive on various open questions of the inflationary
  paradigm. The total number of e-folds of acceleration is given by
  the lifetime of the unstable dust components and is related to the
  duration of reheating. As a result, inflation does not last much
  longer than the number of e-folds of deceleration allowing a small
  spatial curvature and large scale deviations to isotropy to be
  observable today. Energy densities are self-regulated as inflation
  can only start for a total energy density less than a threshold
  value, again related to the species' lifetime. Above this threshold,
  the Universe may bounce thereby avoiding a singularity. Another
  distinctive feature is that the accelerated expansion is of the
  superinflationary kind, namely the first Hubble flow function is
  negative. We show however that the tensor modes are never excited
  and the tensor-to-scalar ratio is always vanishing, independently of
  the energy scale of inflation.}

\maketitle

\section{Introduction}

General Relativity (GR) explains gravitational phenomena in a wide
range of scales, from sub-millimeter to Solar System scales
\cite{Will:2014kxa} and (if we accept the existence of dark matter and
dark energy) even on cosmological scales from Big Bang Nucleosynthesis (BBN)
time until today~\cite{Adam:2015rua}. Despite its observational
success, attempts to modify GR have been proposed in the aim of
alleviating some observational and theoretical problems.

On the one hand, the dark energy problem motivates the search for
modifications of the infrared (IR) regime of gravity in form of
scalar-tensor~\cite{Horndeski:1974wa, Nicolis:2008in, Deffayet:2009wt,
  Deffayet:2009mn, deRham:2011by, Heisenberg:2014kea,
  Heisenberg:2015wja}, vector-tensor~\cite{Jimenez:2008au,Jimenez:2009ai,
  EspositoFarese:2009aj, Jimenez:2009py, BeltranJimenez:2013fca,
  Jimenez:2013qsa,Jimenez:2014rna, Heisenberg:2014rta,
  Tasinato:2014eka}, tensor-tensor~\cite{deRham:2010ik, deRham:2010kj,
  Hassan:2011vm, Hassan:2011ea, Hassan:2011zd, Hinterbichler:2012cn}
and non-local~\cite{Jaccard:2013gla, Ferreira:2013tqn,
  Maggiore:2014sia} theories among others, so that the cosmic
acceleration would be caused by a modification of GR on cosmological
scales rather than induced by the cosmological constant or some new
components~\cite{Copeland:2006wr, Clifton:2011jh}. Also at a
phenomenological level, some models have been put forward to account
for the dark
matter~\cite{Famaey:2011kh,Khoury:2014tka,Blanchet:2015sra,
  Blanchet:2015bia}. From a more theoretical motivation, IR
modifications of gravity have been considered as potential mechanisms
to tackle the old cosmological constant
problem~\cite{Weinberg:1988cp,Martin:2012bt}, like theories exhibiting
degravitating solutions~\cite{Dvali:2002fz, ArkaniHamed:2002fu,
  Dvali:2007kt, deRham:2010tw}, where the cosmological constant is
effectively decoupled from gravity on large scales, or unimodular
gravity, where the additional Weyl symmetry is claimed to prevent
quantum corrections in the form of a cosmological
constant~\cite{Barcelo:2014qva, Carballo-Rubio:2015kaa,
  Alvarez:2015pla} (see however~\cite{Padilla:2014yea, Saltas:2014cta}
for claims in the opposite direction).

On the other hand, the non-renormalizability of GR also calls for
modifications of gravity, in the ultraviolet (UV) regime this time,
which should lead to a gravitational framework compatible with quantum
physics as it is currently understood. These quantum effects are
usually expected to regularise classical singularities present in
Einstein equations. However, modifications of the high curvature
regime have also been explored as possible mechanisms able to resolve
geometric singularities without invoking quantum effects. From a more
phenomenological point of view, very much like IR modifications modify the
late-time cosmology and, therefore, can be used to explain the current
acceleration of the Universe, UV modifications are expected to be at
work in the early Universe. Therefore, one may wonder whether they
could provide a new mechanism for Cosmic
Inflation~\cite{Starobinsky:1979ty, Starobinsky:1980te, Guth:1980zm,
  Linde:1981mu, Albrecht:1982wi, Mukhanov:1985rz, Mukhanov:1988jd}. In
the standard picture, inflation is sourced by a self-gravitating
scalar field in its potential dominated regime. Among the currently
favoured single field models~\cite{Martin:2014vha}, let us notice that
some of them are modified gravity theories, such as the Starobinsky
(or Higgs inflation) model~\cite{Starobinsky:1980te, Bezrukov:2007ep}
which belongs to $f(R)$ theories~\cite{DeFelice:2010aj}.

In this work we explore the possibility of realizing an inflationary
phase through a Born-Infeld modification of gravity in the high
curvature regime. The specific theory that we will consider here
corresponds to the class of modifications introduced
in Ref.~\cite{Jimenez:2014fla}.

The original Born-Infeld theory of electrodynamics was introduced as a
way of regularising the self-energy of point-like charged particles in
electromagnetism~\cite{Born:1934gh}. This was achieved by introducing
a square root structure that gives rise to an upper bound for the
allowed electromagnetic fields. Deser and Gibbons suggested to use the
same idea to resolve the singularities encountered in
GR~\cite{Deser:1998rj}. However, their proposal suffered from an
ambiguity, that originated from the necessity to remove a ghost
present in the metric formulation of the theory. In
Ref.~\cite{Vollick:2003qp}, the theory was considered within the
Palatini formalism, putting forward that, in that approach, the ghost
is naturally avoided.

The phenomenological consequences and viability of Born-Infeld gravity
theories have been extensively explored in cosmology~\cite{Du:2014jka,
  Kim:2013noa,Kruglov:2013qaa, Yang:2013hsa,Avelino:2012ue,
  DeFelice:2012hq, EscamillaRivera:2012vz, Cho:2012vg,
  Scargill:2012kg, EscamillaRivera:2013hv,Banados:2010ix},
astrophysics~\cite{Harko:2013xma, Avelino:2012ge, Sham:2013cya,
  Harko:2013wka, Pani:2011mg}, the problem of cosmic
singularities~\cite{Bouhmadi-Lopez:2013lha, Ferraro:2010at}, black
holes~\cite{Olmo:2015dba, Olmo:2013gqa}, wormhole
physics~\cite{Shaikh:2015oha, Bambi:2015sla, Lobo:2014fma,
  Harko:2013aya} and various extensions of the original formulation
have also been considered~\cite{Odintsov:2014yaa, Makarenko:2014lxa,
  Makarenko:2014nca, Comelli:2005tn, Comelli:2004qr,
  Feigenbaum:1998wy, Feigenbaum:1997pf, Ferraro:2009zk,
  Fiorini:2013kba, Gullu:2010wb, Gullu:2010pc, Gullu:2014gza,
  Nieto:2004qj, Wohlfarth:2003ss, Schmidt-May:2014tpa}.  Another
interesting property of these theories is that they give specific
realizations of Cardassian-like models~\cite{Freese:2002sq,
  Gondolo:2002fh}. Recently, a natural extension of the theory was
introduced in Ref.~\cite{Jimenez:2014fla} where it was shown that the high
curvature regime is free of singularity and may support a quasi-de
Sitter expansion when the Universe is dominated by a perfect fluid
having a vanishing equation of state (see also
Ref.~\cite{Fiorini:2013kba}). UV modifications of GR with additional
degrees of freedom have been shown to provide potential candidates for
dark matter~\cite{Cembranos:2003mr, Cembranos:2003fu, Alcaraz:2002iu,
  Cembranos:2001rp, Cembranos:2004jp, Cembranos:2008gj,
  Jimenez:2014rna} so that finding an UV modification that may also support
accelerated expansion opens the possibility of unifying dark matter and
cosmic inflation.

A notable topic of discussion concerning Palatini theories is the
potential existence of anomalies and surface singularities around
sharp variations of the energy and pressure
densities~\cite{Barausse:2007pn, Pani:2012qd, Kim:2013nna,
  Sham:2013sya, Avelino:2012qe, Sham:2012qi}. As argued in
Ref.~\cite{Pani:2013qfa}, gravitational theories containing auxiliary
fields necessarily introduce new couplings to matter that involve
derivatives of the energy-momentum tensor as source of the ``Einstein
equations''. Thus, systems in which there is a sharp change in the
density profile may lead to strong tidal forces, although
Ref.~\cite{Kim:2013nna} has argued that backreaction effects might
cure such a pathology. We expect these effects to appear for
variations on the density profile of the order of the new mass scale
of the theory, since the corrections with respect to GR are determined
by such a scale. However, in Ref.~\cite{Pani:2012qd} it is argued that
for a specific class of polytropic equations of state, and in a
particular realization of Born-Infeld like theories of gravity, the
appearance of the surface singularity is independent of the scale that
suppresses the corrections with respect to GR. This could also be the
case for the theory that we consider here and addressing this issue
would require a detailed analysis that is beyond the scope of the
present work. We would like to point out, however, that this pathology
is very specific for one theory and one class of equation of state, so
such a problem may be avoided by assuming a high enough scale
suppressing the new corrections. Additionally, when tidal
  forces start growing, additional terms in the fluid description may
  become relevant.

In the following, we consider the Born-Infeld theory introduced in
Ref.~\cite{Jimenez:2014fla} and show that gravitating dust in the high
curvature regime may be used to support Cosmic Inflation. The
inflationary dynamics in the high curvature regime strongly depends on
the speed of sound. In particular, getting enough e-folds of inflation
to solve the usual problems of the Friedmann-Lema\^{\i}tre model,
together with having a graceful exit and a reheating era is a
non-trivial problem. We show that the minimal model satisfying all
these conditions involves a cascade of at least two decaying dust
components ultimately decaying into a radiation fluid. Moreover, using
Big-Bang Nucleosynthesis (BBN) constraints, such a setup gives new
constraints on the energy scale at which the gravity modifications may
take place, which ends up being complementary to the ones coming from
astrophysical processes. The inflationary phase itself exhibits unique
properties as for instance a bounded total number of e-folds, a
maximal energy density and a Hubble parameter which is slowly
increasing during inflation. Although in GR with scalar fields, such a
feature would produce a blue spectral index for the gravitational wave
spectrum, we emphasize here that tensor modes remain unamplified and
inobservable, independently of the energy scale of inflation.

The paper is organized as follows. In section~\ref{sec:theory_mBI} we
give a brief summary of the Born-Infeld inspired modification of
gravity that we will use as well as some of its main properties. Then,
in section~\ref{sec:cosmo_sol} we study isotropic and homogeneous
cosmological solutions taking into account the speed of sound and show
in section~\ref{sec:inflation} how accelerated expansion can be
obtained in the presence of dust. In sections~\ref{sec:twofluids} to
\ref{sec:manyfluids}, we show that a viable model incorporating a
graceful exit together with a reheating before BBN requires the
presence of at least two unstable dust components decaying one into
another and ultimately into radiation. Finally, we discuss various
attractive aspects of the model as well as some properties of the
tensor perturbations in the conclusion.

\section{Minimal Born-Infeld theory}
\label{sec:theory_mBI} 

The minimal extension of the Born-Infeld inspired gravity considered
in the following is described by the action\footnote{Throughout this work, a hat will be used to denote matrix representation of the corresponding tensor.}~\cite{Jimenez:2014fla}
\begin{equation}
\mS=\Mlambda^2\Mp^2\int \d^4x \sqrt{-g}
\Tr\left( \sqrt{\Id+\Mlambda^{-2}\hat{g}^{-1} \hat{R}(\Gamma)}-\Id \right),
\label{eq:gminimal}
\end{equation}
where $\Mp$ is the reduced Planck mass, $\Mlambda$ is some energy
scale (in principle unrelated to $\Mp$), $\gm^{-1}$ ($=g^{\mu\nu}$)
denotes the inverse of the metric tensor, $\Rm$ ($=R_{\mu\nu}$) is the
Ricci tensor matrix, $\Id$ ($=\delta^\mu{}_\nu$) is the $4\times4$
identity matrix and $\Tr(\phantom{x})$ stands for the trace
operator. This action is treated within the Palatini formalism, i.e.,
the Ricci tensor is constructed out of an independent connection field
$\Gamma^\alpha_{\mu\nu}$. The factor in front of the action is chosen
so that we recover the Einstein-Hilbert action at small curvatures and
we subtract the identity inside the square brackets to guarantee the
existence of Minkowski as vacuum solution. This can be seen by
expanding the action at curvatures much smaller than $\Mlambda^2$:
\begin{equation}
\mS=\frac12\Mp^2\int \d^4x\sqrt{-g}
g^{\mu\nu}R_{\mu\nu}(\Gamma)\left[1
  +\order{\frac{\hat{g}^{-1}\hat{R}}{\Mlambda^2}}\right],
\label{eq:gminimal2}
\end{equation}
where we see that the leading order term is nothing but
Einstein-Hilbert action in the Palatini formalism without cosmological
constant.

Varying the action \eqref{eq:gminimal} with respect to the metric
tensor yields the metric field equations
\begin{equation}
(M^{-1})^\alpha{}_{(\mu} R_{\nu)\alpha}-\Tr
  (\Mm-\Id)\Mlambda^2g_{\mu\nu}=\frac{1}{\Mp^2}T_{\mu\nu},
\end{equation}
where we have defined
\begin{equation}
M^\mu_{\phantom{\mu}\nu} \equiv \left( \sqrt{\Id + \Mlambda^{-2}
  \hat{g}^{-1} \hat{R}}\,\right)^\mu_{\phantom{\mu}\nu}.
\label{eq:Mdef}
\end{equation}
Here, we demand that $\Id+\Mlambda^{-2}\gm^{-1}\Rm$ is a positive
definite matrix on physically admissible solutions and we define $\Mm$
as the only positive definite matrix such that
$\Mm^2=\Id+\Mlambda^{-2}\gm^{-1}\Rm$.

The metric field equations can be written in an alternative form by
using the above definition of the fundamental matrix in order to
express the Ricci tensor as $\Rm=\Mlambda^2 \, \gm(\Mm^2-\Id)$. One then obtains
\begin{equation}
\frac12\left[\gm(\Mm-\Mm^{-1}) + (\Mm-\Mm^{-1})^T\gm\right] -
\Tr(\Mm-\Id)\gm = \frac{1}{\Mlambda^2\Mp^2}\Tm,
\label{eq:M}
\end{equation}
where we have used matrix notation and the superscript ``$T$'' stands
for the transposition operator. This equation allows, in principle, to
express the matrix $\Mm$ in terms of the metric tensor and the matter
content by solving an algebraic set of equations. These equations are,
in general, non-linear and several branches may arise in the
theory. Nonetheless, not all of them will be physical since one must
require the matrix $\Mm$ to be positive definite. In addition, there
is only one branch of solutions that will be continuously connected
with GR at low curvatures (or densities). This is indeed the branch
that we will choose in this work, although other branches can also
have interesting phenomenologies \cite{Jimenez:2014fla}.

Variations of the action \eqref{eq:gminimal} with respect to the
connection $\Gamma$ gives the remaining field equations in the Palatini formalism:
\begin{equation}
\nabla_\lambda\left( \sqrt{-g}W^{\beta\nu}\right) -
\delta_\lambda^\beta \nabla_\rho\left(\sqrt{-g}W^{\rho\nu} \right) - 2
\sqrt{-g} \left( \calT_{\kappa \lambda  }^\kappa W^{\beta \nu} -
\delta_\lambda^\beta \calT_{\kappa \rho }^\kappa W^{\rho \nu} +
\calT_{\lambda \rho  }^\beta W^{\rho \nu} \right)=0.
\label{connectionEq}
\end{equation}
in which we have introduced the torsion tensor
$\calT^\mu_{\alpha\beta} \equiv \Gamma^\mu_{[\alpha\beta]}$ and
\begin{equation}
W^{\mu\nu}\equiv (\Mm^{-1})^\mu_{\phantom{\mu}\alpha} g^{\alpha\nu}.
\label{eq:Wdef}
\end{equation}
Assuming all fields to be minimally coupled to the metric, the
connection equations are not sourced by the matter sector, i.e., the
right hand side (RHS) of the connection equations identically
vanishes\footnote{Some subtleties might arise when considering
  fermions, but we will not consider that case here.}. In the
following, we will be interested in torsion-free solutions and therefore set
$\calT^\alpha_{\mu\nu}=0$ from now on. Obviously one needs to check
the consistency of this condition and we show below that it is the
case for the cosmological solutions we are interested in.

For vanishing torsion, taking the trace of \eqref{connectionEq} gives
$\nabla_\lambda\left( \sqrt{-g}\, W^{\lambda \nu}\right)=0$ such that
the connection equations reduce to
\begin{equation}
\nabla_\lambda\left( \sqrt{-g}\, W^{ \beta \nu}\right)=0,
\label{eq:Gamma}
\end{equation}
for all indices. As shown in Ref.~\cite{Jimenez:2014fla}, if the matrix $\Wm$ is
symmetric, an elegant way to solve these equations is to introduce the
auxiliary metric $\gt$ defined by
\begin{equation}
\gt^{\mu\nu} \equiv \sqrt{\det \Mm}g^{\alpha\mu}(\Mm^{-1})^{\nu}{}_\alpha.
\label{eq:gtilde}
\end{equation}
Plugging equations \eqref{eq:Wdef} and \eqref{eq:gtilde} into
\eqref{eq:Gamma} gives
\begin{equation}
\nabla_\lambda\left(\sqrt{-\gt} \gt^{\beta \nu} \right) = 0.
\end{equation}
These equations require $\Gamma$ to be the Levi-Civita connection
associated with the auxiliary metric $\gt$. One can check a posteriori
that having a vanishing torsion is consistent with the solutions where
$W^{\mu\nu}$ is actually symmetric. For a more detailed discussion on
these points see Ref.~\cite{Olmo:2013lta}.

\section{Cosmological solutions}
\label{sec:cosmo_sol}

Homogeneous and isotropic solutions associated with the action
\eqref{eq:gminimal} have been discussed in~\cite{Jimenez:2014fla} for
barotropic fluids and we generalize this approach to any perfect fluid
below.

\subsection{Hubble parameter}

We assume the Friedmann-Lema\^{\i}tre-Robertson-Walker metric
\begin{equation}
\d s^2=-n^2(t)\d t^2+a(t)\d\vec{x}^2,
\end{equation}
where $n(t)$ and $a(t)$ are the lapse and scale factor functions
respectively. The matter sector is modeled by a homogeneous perfect
fluid so that the stress tensor and the fundamental matrix  are assumed to be of the form
\begin{equation}
T_\mu^{\phantom{\mu}\nu}=\diag[-\rho(t), P(t), P(t), P(t)], \qquad
M^\mu_{\phantom{\mu}\nu} = \diag[_0(t), M_1(t), M_1(t), M_1(t)],
\end{equation}
i.e., compatible with the spacetime symmetries. Since the matrix
$\Mm$ must be positive definite, one has $M_0>0$ and $M_1>0$. The
metric field equations (\ref{eq:M}) for the assumed background become
\begin{equation}
\begin{aligned}
\frac{1}{M_0}+3M_1&= 4+\rhob, \\
M_0+2M_1+\frac{1}{M_1}&= 4-\Pb, 
\end{aligned}
\label{eqM0M1}
\end{equation}
where we have defined the dimensionless quantities
\begin{equation}
\rhob \equiv \dfrac{\rho}{\Mlambda^2\Mp^2}\,,\qquad \Pb \equiv \dfrac{P}{\Mlambda^2\Mp^2}\,.
\end{equation}
These equations allow to obtain $M_0(\rho,P)$ and $M_1(\rho,P)$
algebraically so that we can compute the auxiliary metric that generates
the connection in terms of the matter content. According to \eqref{eq:gtilde}, the auxiliary
metric also takes a FLRW form
\begin{equation}
\d \tilde{s}^2=-\nt^2(t)\d t^2+\at^2(t)\d\vec{x}^2,
\end{equation}
where the auxiliary lapse and scale factor functions are given by
\begin{equation}
\tilde{n}^2(t)=  n^2(t)\sqrt{M_0 M_1^{-3}}, \qquad \tilde{a}^2(t)=\frac{a^2(t)}{\sqrt{M_0 M_1}}\,.
\label{HtToH}
\end{equation}

Since the connection $\Gamma^\alpha_{\mu\nu}$ is associated to the auxiliary
metric $\gt_{\mu\nu}$, the corresponding time-time component of its Einstein tensor
is given by the usual expression in terms of the auxiliary Hubble
parameter $\Ht \equiv \dot{\at}/{\at}$, namely
\begin{equation}
G_{00}(\gt) = 3 \Ht^2 = 3 \left[H - \dfrac{1}{4} \dfrac{\ud \ln(M_0 M_1)}{\ud t} \right]^2,
\label{eq:Gfl}
\end{equation}
with $H=\dot{a}/a$ the Hubble parameter associated to the spacetime metric
$g_{\mu\nu}$. On the other hand, we can use the definition of $\Mm$ in ~\eqref{eq:Mdef} to express the Einstein tensor in terms of $\Mm$ from its definition as follows: 
\begin{equation}
\begin{aligned}
\hat{G}(\gt) & \equiv \Rm(\gt)-\frac12\,\hat{\gt} \,
\Tr(\hat{\gt}^{-1} \Rm)  \\
&= \Mlambda^{2} \, \hat{g}\left[\Mm^2-\Id  -\frac12\Mm \Tr\Big(\Mm-\Mm^{-1}\Big)\right].
\end{aligned}
\label{eq:Gmn}
\end{equation}
Equations \eqref{eq:Gfl} and \eqref{eq:Gmn} will lead to the equivalent of the
usual Einstein equations with a modified source term, since the matrix $\Mm$ is an algebraic function of the matter content. For the FLRW
solutions, the right hand side of equation \eqref{eq:Gmn} reads
\begin{equation}
\begin{aligned}
G_{00}(\gt)  = - \dfrac{\Mlambda^2 n^2(t)}{2} \left(M_0^2 - 1 - 3 M_0
M_1 + 3 \dfrac{M_0}{M_1} \right)  = \Mlambda^2 n^2(t) \left[ M_0^2
+ \dfrac{3}{2} \left(\rhob + \Pb\right) M_0 -1 \right],
\end{aligned}
\label{eq:GooMo}
\end{equation}
where we have used the equation~\eqref{eqM0M1} to express
everything in terms of $M_0$ only. Similarly, the auxiliary Hubble
parameter $\Ht$ can be expressed in terms of $M_0$ by noticing that
equation~\eqref{eqM0M1} implies
\begin{equation}
M_0 M_1 = \dfrac{(4 + \rhob)M_0 -1}{3}\,.
\end{equation}
Then, from equation~\eqref{eq:Gfl} and using~\eqref{eqM0M1}, one gets
\begin{equation}
\Ht = H\left\{1 - \dfrac{M_0}{ 4 \left[(4+\rhob)M_0 -1 \right]} \left[
\dfrac{\ud \rhob}{\ud N} + (4+\rhob) \dfrac{\ud \ln M_0}{\ud N} \right] \right\},
\end{equation}
where $N\equiv \ln[a(t)]$ is the "e-fold" time variable and
$M_0[\rhob(N),\Pb(N)]$ stands for the physical solution of
equations~\eqref{eqM0M1}. Expanding now the total derivative of $M_0$
as
\begin{equation}
\dfrac{\ud \ln M_0}{\ud N} = \dfrac{\partial \ln M_0}{\partial \rhob}
\dfrac{\ud \rhob}{\ud N} + \dfrac{\partial \ln M_0}{\partial \Pb}
\dfrac{\ud \Pb}{\ud N} = \dfrac{\ud \rhob}{\ud N}\left(\dfrac{\partial
  \ln M_0}{\partial \rhob} + \cs^2 \dfrac{\partial \ln M_0}{\partial \Pb}
\right),
\end{equation}
where $\cs^2 \equiv \dot{\Pb}/\dot{\rhob}$ is the sound speed, one
obtains
\begin{equation}
\Ht =  H \left\{1 - \dfrac{M_0}{4\left[(4+\rhob)M_0 -1 \right]}
\dfrac{\ud \rhob}{\ud N} \left[ 1 + (4+\rhob) \left(\dfrac{\partial
  \ln M_0}{\partial \rhob} + \cs^2 \dfrac{\partial \ln M_0}{\partial
    \Pb}\right) \right] \right\}.
\label{eq:Htilde}
\end{equation} 
From equations \eqref{eq:GooMo} and
\eqref{eq:Htilde} one finally obtains the modified Friedmann-Lema\^{\i}tre
equation
\begin{equation}
\Hb^2 = \dfrac{1}{3} \dfrac{M_0^2 + \dfrac{3}{2}(\Pb + \rhob) M_0
  -1}{\left \{1 - \dfrac{M_0}{4\left[(4+\rhob)M_0 -1 \right]}
\dfrac{\ud \rhob}{\ud N} \left[ 1 + (4+\rhob) \left(\dfrac{\partial
  \ln M_0}{\partial \rhob} + \cs^2 \dfrac{\partial \ln M_0}{\partial
    \Pb}\right) \right] \right\}^2}\,,
\label{eq:Hubble}
\end{equation}
where $\Hb \equiv H/\Mlambda$ is the dimensionless Hubble parameter in
units of the new scale $\Mlambda$. As previously advertised, our
Born-Infeld inspired gravity theory within the Palatini formalism has
led us to a version of the Friedmann-Lema\^{\i}tre equation where the
matter source is modified. Let us stress again that
$M_0=M_0(\rhob,\Pb)$ so that the RHS of the above equation is also a
function of $\rhob$ and $\Pb$. A remarkable feature that should not be
unnoticed is the appearance of derivatives of $\rhob$ and $\Pb$, which
will play a crucial role in the dynamics of the system, as we show
below. Moreover, this introduces the novel effect that the background
cosmological evolution is not only determined by the equation of state
parameter $w\equiv \Pb/\rhob$ of the fluid as in the standard case,
but also by the sound speed of the fluid $\cs$.

To end this section we stress that the matter sector is assumed to be
minimally coupled to the metric tensor $g_{\mu\nu}$. In that
situation, conservation of the stress-tensor yields the usual equation
\begin{equation}
\dfrac{\ud \rhob}{\ud N} = -3(\Pb + \rhob).
\end{equation}

\subsection{Physical branch}

\begin{figure}
\begin{center}
\includegraphics[width=1.05\wdblefig]{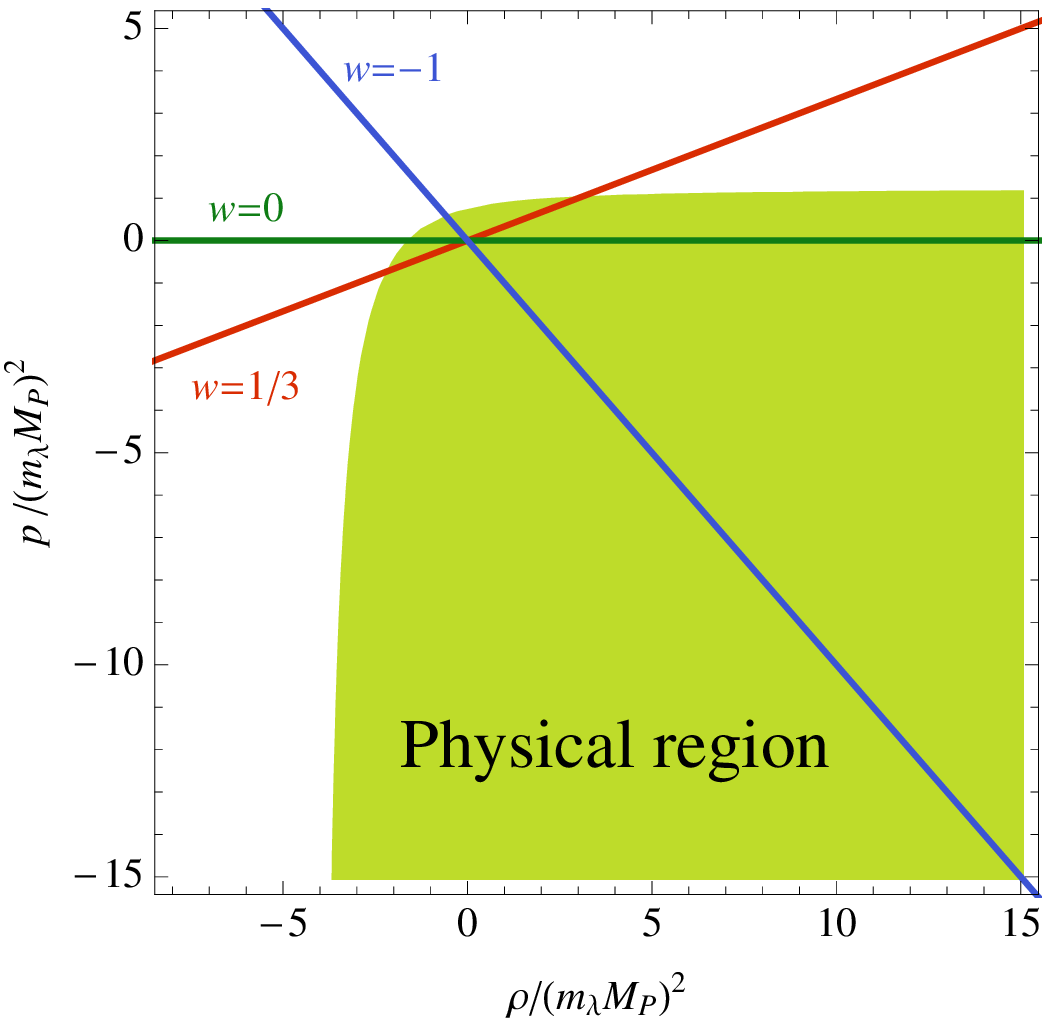}
\includegraphics[width=1.05\wdblefig]{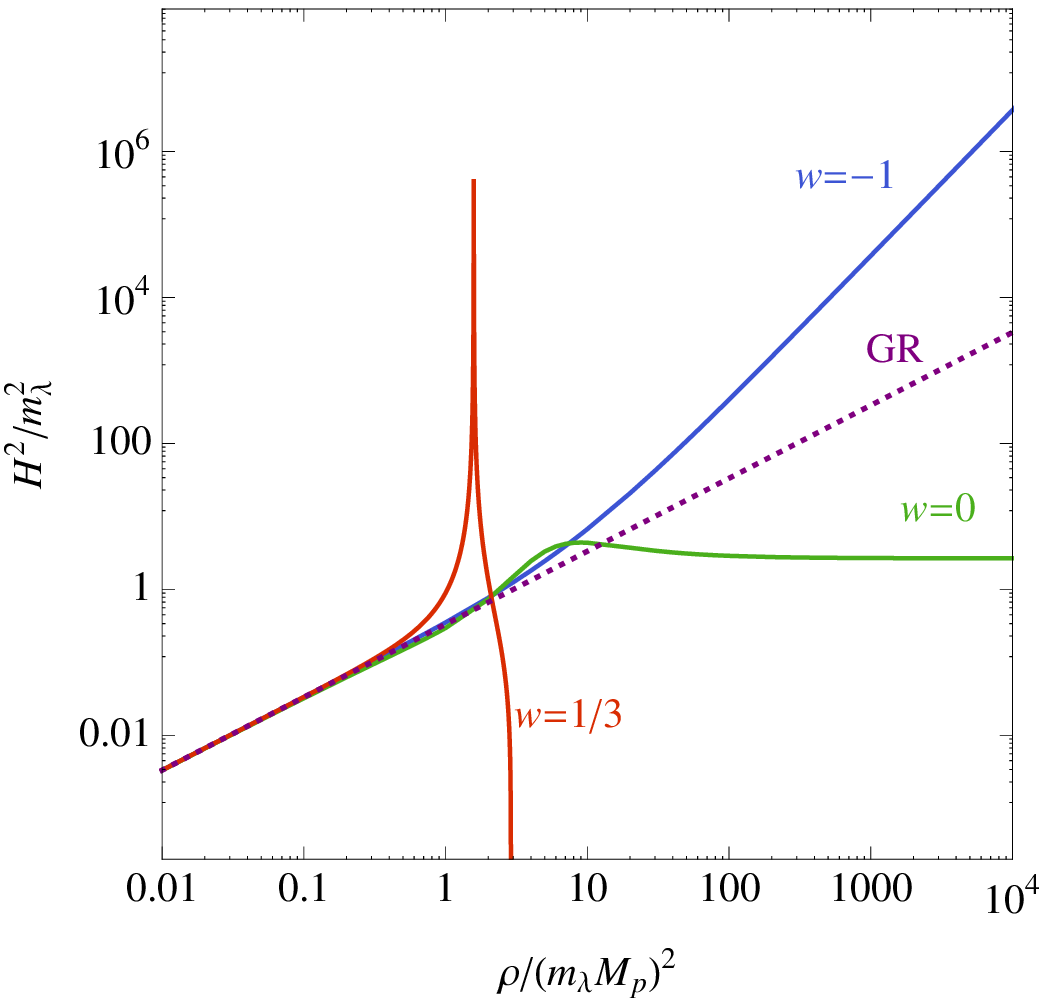}
\caption{Left panel: Region in the plane $(\rhob,\Pb)$ obtained by
  imposing that the fundamental matrix $\Mm$ is positive definite,
  i.e., that both $M_0$ and $M_1$ are real and positive valued. We see
  that such a region is basically defined by the simultaneous
  conditions $\rho \gtrsim -4\Mlambda^2\Mp^2$ and $\Pb \lesssim
  \Mlambda^2\Mp^2$. For an idealized barotropic fluid $P=w \rho$, with
  a constant equation of state parameter $w$, the pressure and/or the
  energy densities may end up being bounded. Right panel: We show the
  Hubble function as a function of the energy density for different
  equation of state parameters illustrating the 3 behaviours discussed
  in the main text. We can see that at low energy densities they all
  follow the usual GR behavior (dotted line), whereas at high energy
  densities ($\rho\gg\Mlambda^2\Mp^2$) the differences appear. In
  particular, for a dust component with $w=0$ we see that $H^2$ goes
  to a constant value, i.e., it gives a de Sitter phase.}
\label{fig:domain}
\end{center}
\end{figure}

The metric field equations \eqref{eqM0M1} generically lead to several
branches for $M_0(\rhob,\Pb)$ and $M_1(\rhob,\Pb)$ due to their
non-linearity. These have been thoroughly discussed in
Ref.~\cite{Jimenez:2014fla} and we simply summarize the results
here. Solving for $M_0$ gives the cubic equation
\begin{equation}
(4+\rhob)M_0^3 +
  \left[\Pb(4+\rhob)+\dfrac{2}{3}(1+\rhob)^2-4\right]M_0^2-\left[\Pb+\dfrac{4}{3}(1+\rhob)\right]M_0
  + \dfrac{2}{3} = 0.
\end{equation}
There are three different branches of solution, but only two of them
are admissible on physical grounds, i.e., obtained by imposing the
positivity of the fundamental matrix $\Mm$. Out of those two physical
solutions, only one branch matches GR at low energy
densities. Although the expression is not particularly illuminating,
it is explicit:
\begin{equation}
\begin{aligned}
M_0 & = \frac{1}{18 (\rhob
   +4)} \left[2^{2/3} \sqrt[3]{27 \sqrt{3} (4+\rhob) \sqrt{-B} - A} +
  \dfrac{2^{1/3} C}{\sqrt[3]{27 \sqrt{3} (4+\rhob)\sqrt{-B} - A}} - D
  \right],
\end{aligned}
\label{eq:Moexplicit}
\end{equation}
where
\begin{equation}
\begin{aligned}
A & = 3456 \Pb^3+\left(108 \Pb^2+576 \Pb+168\right) \rhob ^4-4752
\Pb^2+\left(54 \Pb^3+1080 \Pb^2+1206 \Pb+680\right) \rhob ^3 \\ & +\left(648
\Pb^3+3159 \Pb^2+720 \Pb+2670\right) \rhob ^2 + \left(2592 \Pb^3+1080 \Pb^2+4302
\Pb+2616\right) \rhob \\ & + (72 \Pb+96) \rhob ^5+9144 \Pb+16 \rhob ^6+1456, \\
B & = 144 \Pb^4-1248 \Pb^3+\left(4 \Pb^2-32 \Pb+32\right) \rhob ^4+2404
\Pb^2+\left(9 \Pb^4-468 \Pb^2+384 \Pb+52\right) \rhob ^2 \\ &
+\left(12 \Pb^3-80 \Pb^2-24
\Pb+96\right) \rhob ^3+\left(72 \Pb^4-504 \Pb^3+256 \Pb^2-920
   \Pb+1024\right) \rhob \\ &- 73456 \Pb+1600, \\
C & = \left(18 \Pb^2+144 \Pb+24\right) \rhob ^2+\left(144 \Pb^2+126
\Pb+200\right) \rhob +288 \Pb^2 +(24 \Pb+32) \rhob ^3 +8 \rhob ^4 \\ &- 264
\Pb+488, \\
D & = (6 \Pb+8) \rhob +24 \Pb+4 \rhob ^2-20.
\end{aligned}
\end{equation}
Both $M_0$ and $M_1$ are well-defined, i.e., real and positive, only
on a given domain in the plane $(\rhob,\Pb)$ which has been
represented in Figure~\ref{fig:domain}. 
We have also represented in this figure various barotropic equations
of state $P=w \rho$, with constant $w$, and one can single out three
typical behaviours:
\begin{itemize}
\item For fluids with positive pressure $0<w<1$ we find that there is
  a maximum value for $\rho$ which is given by
  $\rho\lesssim \Mlambda^2M_p^2$. This is the desired property of
  Born-Infeld inspired theories, i.e., we find an upper bound for the
  allowed energy densities.
\item If $0\leq w<-2/3$, there is no upper bound on $\rho$. As
  discussed below, depending on the behaviour of $\cs^2$, the Hubble
  function can take a constant value at high energy densities. Even
  though $\rhob$ can grow indefinitely, the curvature, here parametrically given by
  $H$, remains bounded.
\item Finally, for $-2/3<w\leq -1$ we find that the Hubble function
  grows as $H^2\propto \rho^2$. In this case there is no realization
  of the Born-Infeld mechanism. Interestingly enough, such a behaviour
  is also typical of theories with extra-dimensions~\cite{Cline:1999ts}.
\end{itemize}
As mentioned above, these three ideal cases are only illustrative and
one should keep in mind that in the cosmological context $\cs^2$ is
expected to be a function of time, and thus of the Hubble
parameter. As a result, the trajectory followed by any realistic
gravitating fluid in the plane $(\rhob,\Pb)$ of
Figure~\ref{fig:domain} is in general a curve becoming strongly
non-linear as soon as $\rhob$ or $\Pb$ becomes of order unity.

On the contrary, in the low energy and pressure limit, plugging the
expression for $M_0$ back into the Hubble parameter \eqref{eq:Hubble}
and expanding everything for small $\rhob$ and $\Pb$, one gets
\begin{equation}
\Hb^2 = \dfrac{1}{3} \rhob + \dfrac{2\cs^2-1}{4} \Pb \rhob + \dfrac{4
  \cs^2-1}{8} \rhob^2 - \dfrac{1}{8} \Pb^2 + \order{\rhob^3,\Pb^3}.
\end{equation}
As expected, this expression matches the usual
Friedmann-Lema\^{\i}tre equation for $\rho \ll \Mlambda^2 \Mp^2$ and
$P \ll \Mlambda^2\Mp^2$. As a result, the energy scale at which one
should expect deviations from the standard GR case is determined by
the geometrical mean of $\Mlambda$ and $\Mp$.

\section{Inflationary scenario}
\label{sec:inflation}

In the previous section we have reviewed the cosmological evolution
for the Born-Infeld inspired gravity theory under consideration. We
have shown that for a perfect fluid with equation of state parameter
satisfying $-2/3<w\leq0$, the energy density may not be bounded from
above, but the curvature is. In particular, as we show in more detail
in the next section, for a dust gravitating fluid having $w=0$, the
Hubble parameter becomes nearly constant at high energy densities
thereby allowing a quasi-de Sitter expansion typical of an
inflationary epoch~\cite{Mukhanov:1990me}. Since the theory matches GR
at low energies, the inflationary graceful exit is always naturally
realised within our Born-Infeld inspired gravity when such a regime is
reached. However, with only dust, the Universe would end up being
matter dominated after inflation and, in order to produce a radiation
dominated Universe, one needs to implement some mechanism allowing the
dust to decay into radiation. This is analogous to the reheating
period within the standard inflationary picture where the inflaton
decays at the end of inflation giving rise to radiation made out of
relativistic degrees of freedom.

Thus, in section~\ref{sec:twofluids}, instead of stable dust, we will
consider an unstable dust fluid decaying into radiation. However, in
this scenario the sound speed is no longer vanishing, see
equation~\eqref{eq:Hubble}, and this has a crucial effect since now
the existence of a quasi-de Sitter era becomes possible only within a
finite range of energy densities. Although inflation can be made to
last long enough by adequately fixing the decay rate of the dust
component, we show that the duration of the reheating era ends up
being necessarily longer than the inflationary period. On the other
hand, the decay rate will also determine the beginning of the
radiation era, so the duration of inflation and the reheating period
are linked. This in fact prevents the model from solving the flatness
problem of the standard FLRW cosmology.

In section~\ref{sec:manyfluids}, this issue is solved by considering a
cascade of decaying dust fluids which ultimately end into
radiation. In such a case, inflation can be realized with a graceful
exit onto a reheating era. The solution comes about because while the
duration of inflation depends on the whole cascade of dust fluids, the
reheating duration depends only on the decay rate of the last
component thereby making the model viable.

\subsection{Stable dust}

\begin{figure}
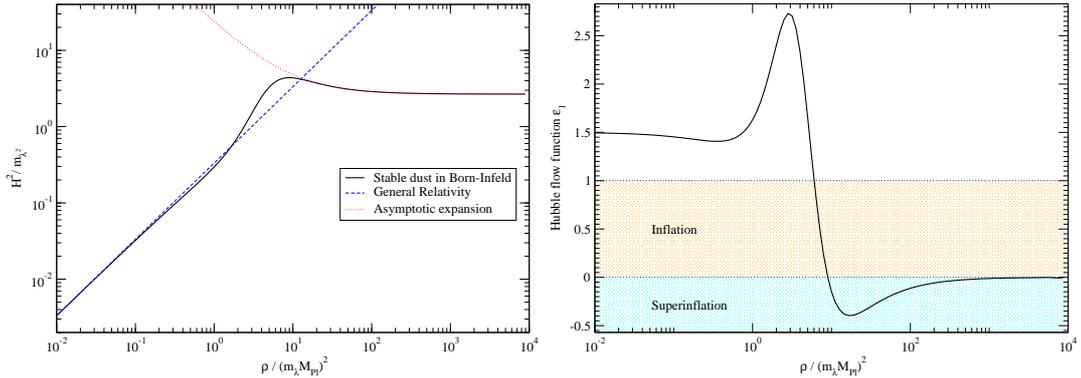

\begin{center}
\includegraphics[width=1.01\wdblefig]{stabledust}
\includegraphics[width=\wdblefig]{epsilon}
\caption{Hubble parameter square $\Hb^2$ as a function of the stable
  dust energy density $\rhob$ (left panel). In the Born-Infeld regime,
  $\rhob \gg1$, the Hubble parameter is nearly constant and slowly
  grows $\dot{H}>0$ to generate a superinflationary era. The red
  dotted curve corresponds to the asymptotic expansion of
  equation~\eqref{eq:HdustAsymp}. In the right panel, the first Hubble flow
  function $\epsilon_1$ is represented as a function of $\rhob$. In
  the superinflationary regime $\epsilon_1<0$.}
\label{fig:Hdust}
\end{center}
\end{figure}

Let us start by considering the gravitating fluid to be pure dust and
conserved characterized by
\begin{equation}
w=\dfrac{\Pb}{\rhob} = 0, \qquad \cs^2 = 0, \qquad \dfrac{\ud \rhob}{\ud N} = -3 \rhob.
\end{equation}
Plugging these conditions into equations~\eqref{eq:Hubble} and
\eqref{eq:Moexplicit} gives the Hubble parameter as an algebraic
function of $\rhob$ only. If we take the high energy density limit
$\rhob \gg 1$ (i.e., in the Born-Infeld regime), we find
\begin{equation}
\Hb = \sqrt{\dfrac{8}{3}} + \dfrac{8 \sqrt{3} - 3\sqrt{6}}{\rhob} + \order{\dfrac{1}{\rhob^2}},
\label{eq:HdustAsymp}
\end{equation}
so that the Hubble parameter becomes nearly constant and the
expansion of the Universe is accelerated. The exact dependence of
$\Hb$ with respect to $\rhob$ as well as the above expansion are
represented in Figure~\ref{fig:Hdust}. Strictly speaking, acceleration
of the scale factor, i.e. inflation, occurs as long as the first Hubble flow
function $\epsilon_1 \equiv - \ud \ln(H) /\ud N$ is less than
unity. From equation~\eqref{eq:HdustAsymp} we obtain
\begin{equation}
\epsilon_1 = -\dfrac{9 (4\sqrt{2}-3)}{2 \rhob} + \order{\dfrac{1}{\rhob^2}},
\label{eq:eps1}
\end{equation}
such that for $\rhob \gg1$ one has $\epsilon_1 \lesssim 0$. Negative
values of $\epsilon_1$ cannot be obtained within General Relativity
and correspond to superinflation~\cite{Gunzig:2000kk}. Given that the
gravitational sector is not described by the usual Einstein-Hilbert
term, let us emphasize that it is not possible to deduce from
$\epsilon_1 < 0$ that the primordial perturbations will have a nearly
scale invariant power spectrum with a blue spectral index. The
cosmological perturbations are discussed in more detail in the
conclusion where we show that tensor perturbations are in fact not
generated.

Finally, as can be seen in Figure~\ref{fig:Hdust}, as soon as $\rhob
\lesssim 1$, the Hubble parameter evolution becomes nearly identical
to GR (up to some relaxation oscillations) and inflation naturally
ends. However, in the subsequent decelerated expansion, the Universe
is and will remain matter dominated, which is in contradiction with
both BBN and the existence of the Cosmic Microwave Background
(CMB). The most natural extension to produce a radiation dominated
Universe after inflation is to assume that the dust component is actually
decaying into radiation. We explore this possibility in the next section.

\subsection{Decaying dust and radiation}
\label{sec:twofluids}

In order to generate a radiation dominated universe after inflation, let us now consider an unstable dust component that decays into radiation
at a constant rate $\Gamma$. Thus, the
matter sector consists of two fluids, decaying dust and radiation, in
interaction, which are described by the following coupled equations:
\begin{equation}
\dfrac{\ud \rhob_1}{\ud t} + 3 H \rhob_1 = -\Gamma \rhob_1, \qquad
\dfrac{\ud \rhobr}{\ud t} + 4 H \rhobr = \Gamma \rhob_1,
\label{eq:toytwo}
\end{equation}
where $\rhob_1$ and $\rhobr$ are the energy densities of dust and
radiation respectively and $H$ is given by
equation~\eqref{eq:Hubble}. Again, since the matter sector is
minimally coupled, we have conservation of the total energy density
$\rhob = \rhobr + \rhob_1$. However, the total pressure does no longer
vanish and reads $\Pb = \rhobr/3$. In addition, we are also in the
presence of a time-dependent sound speed $\cs^2$.

We can rewrite the equations of the interacting dust and radiation
system in a more convenient manner by introducing the radiation
fraction
\begin{equation}
X \equiv \dfrac{\rhobr}{\rhob} = \dfrac{\rhobr}{\rhobr + \rhob_1}\,,
\end{equation}
so that equation~\eqref{eq:toytwo} can be recast into equations of evolution
for $\rhob$ and $X$
\begin{equation}
\dfrac{\ud \rhob}{\ud N} = - (3 + X) \rhob, \qquad \dfrac{\ud X}{\ud
  N} = (X-1) X + (1-X) \dfrac{\Gammab}{\Hb}\,,
\label{eq:evoltwo}
\end{equation}
where $\Gammab \equiv \Gamma/\Mlambda$ and $\Pb = X \rhob/3$. From the
definition of $\cs^2 \equiv \dot{\Pb}/\dot{\rhob}$, one gets
\begin{equation}
\cs^2 = \dfrac{4X}{3(3+X)} - \dfrac{1-X}{3(3+X)} \dfrac{\Gammab}{\Hb}\,.
\label{eq:cstwo}
\end{equation}
Because the above expression explicitly involves $\Hb$, and
equation~\eqref{eq:Hubble} depends on $\cs^2$, the resulting evolution
of the system is highly non-linear and crucially depends on the
functional form of $\cs^2$. As discussed before, because the speed of
sound ``gravitates'' within Palatini theories, this is expected and
emphasizes the importance of considering more than the equation of
state parameter in the fluid description in order to obtain the
background cosmological evolution.

Plugging equation~\eqref{eq:cstwo} into equation~\eqref{eq:Hubble}
yields an algebraic equation for $\Hb$ that can be analytically solved
in terms of $\Pb$ and $\rhob$ as\footnote{When solving this equation
  we obtain two branches. We choose the one corresponding to an
  expanding rather than contracting universe.}
\begin{equation}
\Hb = \dfrac{\sqrt{\dfrac{M_0^2}{3} + \dfrac{1}{2}(\Pb + \rhob) M_0
    -\dfrac{1}{3}} + \dfrac{3(\Pb+\rhob)}{4} \dfrac{(4+\rhob) \,
    \csb^2 \, \Gammab }{(4+\rhob) M_0 -1} \dfrac{\partial \ln
    M_0}{\partial \Pb}}{1 + \dfrac{3(\Pb + \rhob)}{4}
  \dfrac{M_0}{(4+\rhob)M_0-1} \left[1 + (4+\rhob)\left(\dfrac{\partial
      \ln M_0}{\partial \rhob} + \csa^2 \dfrac{\partial \ln
      M_0}{\partial \Pb} \right) \right]}\,,
\label{eq:HubbleMix}
\end{equation}
where we have defined
\begin{equation}
\csa^2 \equiv \dfrac{4X}{3(3+X)}\,,  \qquad \csb^2 \equiv \dfrac{1-X}{3(3+X)}\,.
\end{equation}
This expression matches equation~\eqref{eq:Hubble} if one sets both
the decay rate $\Gammab$ and the radiation fraction $X$ to zero, as
one may expect. However, for non-vanishing $\Gammab$, the limit $X
\rightarrow 0$ does not give back the pure dust behaviour as there
will always be a term proportional to $\Gammab$ in the expression of
$\Hb$. This is important for the initial conditions of the
inflationary scenario since it will set a maximum value for the total
energy density. More precisely, expanding
equation~\eqref{eq:HubbleMix} at large $\rhob$ and small $X$ we obtain
for the initial value of the Hubble parameter
\begin{equation}
\begin{aligned}
\Hb_{\mathrm {ini}} &= \sqrt{\dfrac{8}{3}} - \dfrac{21+\sqrt{2}}{24} \, \Gammab
-\dfrac{\Gammab}{6\sqrt{2}} \, \rhob + \left(\dfrac{4}{3 \sqrt{3}} -
\dfrac{4+149\sqrt{2}}{288} \, \Gammab \right)X \rhob -
\dfrac{4+\sqrt{2}}{72}\, \Gammab \, X \rhob^2 \\ & +
\left(\dfrac{-124+189\sqrt{2}}{12 \sqrt{3}} + \dfrac{-8548 + 4217
  \sqrt{2}}{1152}\, \Gammab \right) X + \left( 8 \sqrt{3} - 3 \sqrt{6}
+ \dfrac{248 - 249 \sqrt{2}}{64} \, \Gammab \right) \dfrac{1}{\rhob}
\\ & + \left(\dfrac{829}{3 \sqrt{3}} - 59 \sqrt{6} + \dfrac{454560 -
  292333 \sqrt{2}}{4608} \, \Gammab \right) \dfrac{X}{\rhob} +
\order{\dfrac{1}{\rhob^2},X^2}.
\label{eq:HmixAsymp}
\end{aligned}
\end{equation}

As opposed to pure stable dust, there is now a maximal value of
$\rhob$ at which the Hubble parameter vanishes. Such a behaviour
happens for any value of $X$, even for $X=0$, and therefore is not due
to non-vanishing pressure but induced by a non-vanishing speed of
sound. Let us stress that this is only relevant for the initial
conditions since, if we have a non-vanishing value of $\Gamma$, the
evolution of the system will immediately generate a certain amount of
radiation. Under these considerations and at leading order in small
$X$ and $\Gammab$, one gets $\Hb(\rhob_{\max}) = 0$ for
\begin{equation}
\rhob_{\max} =\sqrt{\frac83}\left(\frac{\Gammab}{6\sqrt{2}}-\frac{4}{3\sqrt{3}}X\right)
\label{eq:rhomaxX}
\end{equation}
where we see that for $X=0$, we obtain the non-vanishing maximum value
\begin{equation}
\rhob_{\max} = \dfrac{24}{\sqrt{3} \Gammab}\,.
\label{eq:rhomax}
\end{equation}
This condition guarantees that the universe starts in an expanding
phase, so the inflationary period will be achieved. We should
emphasize that this bound appears if we impose a vanishing radiation
component initially. If we have some radiation initially, then
\eqref{eq:rhomaxX} should be used instead. Interestingly,
  initial energy densities higher than $\rhob_{\max}$ continuously map
  into negative values of the initial Hubble parameter. This suggests
  that some of our inflationary solutions could be extended back in
  time with a bounce~\cite{Falciano:2008gt, Peter:2008qz,
    Lilley:2011ag}.

In order to discuss the inflationary phase only, we will assume in the
following that the Universe starts its evolution with $\rhob(0) =
\rhobini \lesssim \rhob_{\max}$, in the regime in which $\Hb$ is
independent of $\rhob$, and with only decaying dust, i.e. having
$X(0)=0$. Equations~\eqref{eq:evoltwo} can be approximated by
\begin{equation}
\dfrac{\ud X}{\ud N} \simeq -X + \dfrac{\Gammab}{\Hb}\,, \qquad
\dfrac{\ud \rhob}{\ud N} \simeq -3 \rhob,
\end{equation}
the solutions of which are
\begin{equation}
X(N) \simeq \dfrac{\Gammab}{\Hb}\,, \qquad \rhob(N) = \rhobini \,  e^{-3N}.
\end{equation}
Therefore, while the Universe is inflating, the radiation fraction
remains constant and given by $\Gammab/\Hb$. This in turn allows to
obtain the condition guaranteeing that the system remains inside the
physical region of the theory as depicted in Figure
\ref{fig:domain}. The radiation component gives rise to a positive
pressure given by $\Pb\simeq\frac13 X\rhob$ so that, according to the
above solution for $X$, we have $\Pb\simeq \Hb\rhob/(3\Gammab)$ so the
condition to be inside the physical region $\Pb\lesssim 1$ leads to a
bound for $\rhob$ given by $\rhob\lesssim
\sqrt{8}/(3\sqrt{3}\Gammab)$. It is of the same order of magnitude as
the bound given by \eqref{eq:rhomax}, although obtained from a
different condition.

On the other hand, the total energy density is driven by the dust
component and decreases exponentially fast. Inflation ends for
$\epsilon_1(\rhob)=1$, i.e. when we exit the Born-Infeld regime and
the evolution equations become close to those in GR. In order to
obtain the duration of the inflationary era, we only need an order of
magnitude estimate for $\rhob(\N{end})$ which can be obtained by
solving $\Hb(\rhob) = \sqrt{\rhob/3}$ by using the asymptotic
expansion~\eqref{eq:HmixAsymp}.  At leading order, one gets
$\rhob(\N{end}) \simeq 8$ and, therefore, the maximum number of
e-folds of inflation is
\begin{equation}
\Delta \N{inf} = \dfrac{1}{3} \ln \left(\dfrac{\rhob_{\max}}{8}
\right) \simeq  \dfrac{1}{3} \ln \left(\dfrac{\sqrt{3}}{\Gammab} \right).
\end{equation}
In order to have more than $60$ e-folds of inflation (which is the
typically required minimum duration of inflation), the decay rate of
the dust component should be very small, namely $\Gamma < 10^{-78}
\Mlambda$, which compromises the appearance of the radiation era after
inflation. For $\Mlambda$ of the order of the Planck mass, this
implies that the dust component should have a lifetime around
$10^{34}\,\us$, i.e. much longer than the (GR) age of the Universe
$\simeq 10^{17}\,\us$.

Considering a mass scale $\Mlambda$ much larger than the Planck mass
does not help either. Indeed, the reheating proceeds only when the
cosmological evolution matches GR, and the radiation era starts when
the reheating is completed at $X(\N{reh}) \simeq 1$. Assuming $X \ll
1$ during reheating, we can approximate the evolution equations by
\begin{equation}
\dfrac{\ud X}{\ud N} \simeq -X + \dfrac{\sqrt{3} \,
  \Gammab}{\rhob^{1/2}}, \qquad \rhob(N>\N{end}) \simeq \rhob(\N{end}) \,
e^{-3 \Delta N} \simeq 8 e^{-3 \Delta N},
\end{equation}
where $\Delta N \equiv N - \N{end}$. Therefore, with $X(\N{end})
\simeq 0$, one gets
\begin{equation}
X(N>\N{end}) \simeq e^{-\Delta N} \int_{\N{end}}^N e^{\Delta n}
\dfrac{\sqrt{3} \Gammab}{\rho^{1/2}(n)}\, \ud n \simeq
\dfrac{\sqrt{3}\,\Gammab}{5\sqrt{2}} e^{3 \Delta N/2},
\end{equation}
and solving for $X\simeq 1$ gives
\begin{equation}
\Delta \N{reh} \simeq \dfrac{2}{3}  \ln\left(\dfrac{5
  \sqrt{2}}{\sqrt{3}\,\Gammab} \right) = 2 \Delta \N{inf} +
\dfrac{1}{3} \ln \left(\dfrac{50}{9} \right).
\end{equation}
One concludes that, independently of $\Gammab$ and $\Mlambda$, the
reheating era typically lasts twice the number of e-folds of
inflation. Because reheating is a decelerating era, this
catastrophically prevents the inflationary era to solve the flatness
and horizon problem. The problem is that the decay rate determines both the duration of inflation and the reheating period so that both are intimately linked. In the next section we avoid this difficulty by introducing more than one decaying dust component.

\subsection{Cascading dust and radiation}
\label{sec:manyfluids}

In order to achieve a viable inflationary scenario leading to a
radiation dominated universe, one is led to consider a cascade of
unstable dust components decaying one into another and ultimately
producing radiation. Intuitively, the number of e-folds required to
reheat the Universe will be set by the lifetime of the most stable
specie whereas the speed of sound evolution, and thus inflation, is
expected to be sensitive to all species or, equivalently, all the
unstable components will contribute to generating a radiation fraction
and, thus, to the total pressure. For this scenario, one finds that
the relation between $\Delta \N{reh}$ and $\Delta \N{inf}$ is indeed
relaxed. The conservation equations for each species are given by
\begin{equation}
\begin{aligned}
\dfrac{\ud \rhob_i}{\ud t} +3H \rhob_i &= \Gamma_{i-1}\rhob_{i-1}-\Gamma_i\rhob_i \qquad i=1,...,n\\
\dfrac{\ud \rhobr}{\ud t} + 4H \rhobr &= \Gamma_{n} \rhob_{n},
\end{aligned}
\end{equation}
with $\Gamma_0=\rhob_0=0$ and where $\rhob_i$ are the energy densities
of the dust components, $\Gamma_i$ are the decay rates of the
corresponding particles and $\rhobr$ is the energy density of the
radiation component as before. Introducing the relative fractions
\begin{equation}
X_i \equiv \dfrac{\rhob_i}{\sum_{i=1}^n \rhob_i + \rhobr}\,, \qquad X=
\dfrac{\rhobr}{\sum_{i=1}^n \rhob_i + \rhobr}\,,
\end{equation}
one has now to solve the coupled system of equations
\begin{equation}
\dfrac{\ud X_i}{\ud N} = \left(X - \dfrac{\Gammab_i}{\Hb}
\right) X_i + \dfrac{\Gammab_{i-1}}{\Hb} X_{i-1}\,, \qquad \dfrac{\ud
  X}{\ud N} = (X-1)X + \dfrac{\Gammab_n}{\Hb} X_n,
\label{eq:evolmulti}
\end{equation}
with $\Hb$ given in equation~\eqref{eq:Hubble}. Again, the total
energy density $\rhob$ is conserved so that it satisfies
~\eqref{eq:evoltwo}. The total pressure is still driven by the
radiation component and reads $P=X \rhob/3$ whereas the sound speed
becomes
\begin{equation}
\cs^2 = \dfrac{4X}{3(3+X)} - \dfrac{X_n}{3(3+X)} \dfrac{\Gammab_n}{\Hb}\,.
\end{equation}

\subsubsection{Inflationary regime}

Comparing this expression to equation~\eqref{eq:cstwo}, one deduces that the
Hubble parameter for cascading dust and radiation is given by
equation~\eqref{eq:HubbleMix} with
\begin{equation}
\csa^2 \equiv \dfrac{4X}{3(3+X)}\,, \qquad \csb^2 \equiv \dfrac{X_n}{3(3+X)}\,.
\end{equation}
As a result, the Hubble parameter has an explicit dependence only on
the total energy density $\rhob$, the radiation fraction $X$, the
fraction $X_n$ of the last specie in the decaying cascade as well as its
associated decay rate $\Gammab_n$ (into radiation). Expanding the
Hubble parameter at large value of $\rhob$ and small values of $X$ and
$X_n$, one gets
\begin{equation}
\begin{aligned}
& \Hb  = \sqrt{\dfrac{8}{3}} - \dfrac{21+\sqrt{2}}{24} \, \Gammab_n X_n -
\dfrac{\Gammab_n X_n}{6\sqrt{2}}\, \rhob + \left(\dfrac{4}{3\sqrt{3}}
- \dfrac{4+173 \sqrt{2}}{288} \Gammab_n X_n \right) X \rhob -\dfrac{4
  + \sqrt{2}}{72} \Gammab_n X_n X \rhob^2 \\
& + \left(\dfrac{-124 + 189 \sqrt{2}}{12 \sqrt{3}} + \dfrac{-9556 +
  4169 \sqrt{2}}{1152} \Gammab_n X_n \right) X + \left( 8 \sqrt{3} - 3 \sqrt{6}
+ \dfrac{248 - 249 \sqrt{2}}{64} \, \Gammab_n X_n \right) \dfrac{1}{\rhob}
\\ & + \left[ \dfrac{\sqrt{3}\left(856 + 493 \sqrt{2}\right)}{90 + 63 \sqrt{2}} +
  \dfrac{7\left(67488 - 44323 \sqrt{2}\right)}{4608} \Gammab_n X_n \right]
  \dfrac{X}{\rhob} + \order{\dfrac{1}{\rhob^2},X^2,X_n^2}.
\end{aligned}
\label{eq:HubbleMulti}
\end{equation}
Up to some numerical coefficients, this expression is formally
identical to equation~\eqref{eq:HubbleMix} with the replacement
$\Gammab \rightarrow \Gammab_n X_n$. Therefore, starting within the
superinflationary regime at vanishing radiation now requires $\rhob <
\rhob_{\max}$ with
\begin{equation}
\rhob_{\max} \simeq  \dfrac{24}{\sqrt{3} \Gammab_n X_n}\,,
\label{eq:rhomaxXn}
\end{equation}
and the duration of inflation can, a priori, be made longer by having
small values of $X_n$ instead of $\Gammab_n$. However, $X_n$ is not
arbitrary but given by solving the set of
equations~\eqref{eq:evolmulti}. The crucial point to realize here is
that, in the previous case with only one dust component, having small
$X$ required having small $\Gamma$ as well, i.e., the two conditions
were related, whereas in the cascading case the condition will depend
on the whole set of decay rates, as we show in the following.

Assuming $\Hb$ to be almost constant with $\rhob(0) = \rhobini
\lesssim \rhob_{\max}$ together with $X_1(0)=1$, $X_{i \ne 1}(0)=0$ and
$X(0)=0$, one can find an approximate solution for all the
species. One gets
\begin{equation}
\dfrac{\ud X_1}{\ud N} \simeq - \dfrac{\Gammab_1}{\Hb} X_1, \qquad
\dfrac{\ud X_i}{\ud N} \simeq -\dfrac{\Gammab_i}{\Hb}X_i +
\dfrac{\Gammab_{i-1}}{\Hb} X_{i-1}.
\label{eq:evolmultiapprox}
\end{equation}
The first dust component can be immediately integrated as
\begin{equation}
X_1(N) = \exp\left(-\dfrac{\Gammab_1}{\Hb}
N\right) \simeq 1,
\end{equation}
where the last approximation requires $N \ll \Hb/\Gammab_1$. For
instance, if we require the duration of inflation to be
$\order{10^2}$, we need $\Gammab_1\lesssim 10^2\Hb$. Obviously this is
nothing but the condition that the dust component should remain stable
throughout the whole inflationary phase. The remaining components of
the cascade $X_i$'s are given by the quadrature
\begin{equation}
X_i(N) \simeq \dfrac{\Gammab_{i-1}}{\Hb} \int_0^N e^{(\Gammab_i/\Hb)
  (y-N)} X_{i-1}(y) \ud y.
\end{equation}
Plugging $X_1 \simeq 1$ into this equation for $i=2$ gives
\begin{equation}
X_2 (N) \simeq \dfrac{\Gammab_1}{\Gammab_2} \left[1 -
e^{-(\Gammab_2/\Hb) N} \right] \simeq \dfrac{\Gammab_1}{\Hb} N\,,
\end{equation}
where the last approximation is valid for $N \ll
\Hb/\max(\Gammab_1,\Gammab_2)$ (which is again related to the stability condition of the cascade during inflation). From this expression for $X_2(N)$ one
can compute $X_3(N)$ and so on such that the full hierarchy is given by
\begin{equation}
X_n(N) \simeq \dfrac{N^{n-1}}{(n-1)!}\prod_{j=1}^{n-1} \dfrac{\Gammab_j}{\Hb}\,,
\label{eq:Xninf}
\end{equation}
again for $N \ll \Hb/\max(\{\Gammab_i\}_{i=1,{n-1}})$. As a result,
concerning the inflationary phase, the cascade of dust components
behave as one unstable dust fluid with an effective decay rate given
by
\begin{equation}
\Gammab = \Gammab_n X_n \simeq \dfrac{\prod_{i=1}^n
  \Gammab_i}{\Hb^{n-1}} \simeq \dfrac{1}{(n-1)!}\left(\dfrac{3 N^2}{8} \right)^{\frac{n-1}{2}}
\prod_{i=1}^n \Gammab_i\,.
\end{equation}
Although $\Gammab$ is no longer constant, its dependence on the e-fold
number $N$ is only a power law such that $\rhob_{\max}$ acquires the
same dependence on $N$. Since the total energy density $\rhob(N)$
decreases much faster, as $\exp(-3N)$, it is only necessary to require
$\rhob < \rhob_{\max}$ at the beginning of inflation. In order to show
this more precisely, we need to impose $\rhob/\rhob_{\max}\lesssim 1$
throughout the whole inflationary period. Then, we only need to
compute the number of e-fold at which $\rhob/\rhob_{\max}\lesssim $
takes its maximum value\footnote{As in the previous cases, this is
  quantitatively similar to imposing the total pressure to be smaller
  than $\Mlambda^2\Mp^2$, i.e., $\Pb=1/3 X \rhob=1/3 \Gammab_n
  X_n\rhob\lesssim 1.$}. We find that this happens for
$N=(n-1)/3$. Since we expect inflation to last for about 60 e-folds,
the obtained value will only make sense if it is smaller than $\sim
60$. This is indeed the case if $n-1<180$, so we will assume this in
the following. Thus, taking $N=(n-1)/3$ in the previous expression
gives the maximal number of inflationary e-folding
\begin{equation}
\Delta \N{inf} \simeq \dfrac{1}{3}
\ln\left(\dfrac{\sqrt{3}}{\prod_{i=1}^n \Gammab_i} \right) +
\dfrac{n-1}{6} \ln\left[\dfrac{24}{(n-1)^2} \right] + \dfrac{1}{3} \ln [(n-1)!].
\label{eq:Ninf}
\end{equation}
For instance, taking all the $\Gammab_i$ equals, getting $60$ e-folds
of inflation requires $\Gamma_i < 10^{-26} \Mlambda$ for a mixture of
$n=3$ dust components. Taking $\Mlambda=\order{\Mp}$, the minimum
lifetime of each specie is required to be greater than
$10^{-19}\,\us$.

\subsubsection{Reheating}

The reheating can be dealt as in the previous section. Provided the
radiation component remains subdominant, $X \ll 1$, one has $\Hb
\simeq \sqrt{\rhob/3}$ with $\rhob(N) \simeq \rhob(\N{end}) \exp(-3 \Delta
N)$. Plugging these expressions into
equation~\eqref{eq:evolmultiapprox} gives the evolution of the density
fraction of all the species during reheating, up to the end at which
the radiation fraction becomes significant. For $X_1$, one gets
\begin{equation}
X_1(N>\N{end}) = X_1(\N{end}) \exp\left[-\dfrac{2}{3}
\dfrac{\Gammab_1}{\Hbend}\left(e^{3 \Delta N/2} -1 \right) \right],
\label{eq:X1reh}
\end{equation}
where $\Hbend = \sqrt{\rhob(\N{end})/3} \simeq \sqrt{8/3}$. Moreover,
under the reasonable assumption that $X_1$ does not evolve much during
inflation, i.e. $\Delta \N{inf} \ll \Gammab_1^{-1} \sqrt{8/3} $, one has
$X_1(\N{end}) \simeq 1$. During reheating, $X_1(N)$ remains almost
constant as long as $\Delta N \ll \Delta N_1$ where
\begin{equation}
\Delta N_i \equiv N_i - \N{end} = \dfrac{2}{3} \ln \left(\dfrac{3}{2}
\dfrac{\Hbend}{\Gammab_i} \right) \simeq \dfrac{2}{3} \ln
\left(\dfrac{\sqrt{6}}{\Gammab_i} \right).
\end{equation}
For the other species, one obtains for $i>1$
\begin{equation}
\begin{aligned}
X_i(N) & \simeq X_i(\N{end})  \exp\left[-\dfrac{2}{3}
\dfrac{\Gammab_i}{\Hbend}\left(e^{3 \Delta N/2} -1 \right) \right]
\\ & + \dfrac{\Gammab_{i-1}}{\Hbend} 
\int_{\N{end}}^N \exp{\left[\dfrac{2}{3}
  \dfrac{\Gammab_i}{\Hbend} \left( e^{3 \Delta y/2} -e^{3 \Delta N/2} \right) +
  \dfrac{3}{2} \Delta y \right]} X_{i-1}(y) \, \ud y.
\end{aligned}
\label{eq:Xireh}
\end{equation}
From the inflationary solution, $X_i(\N{end})$ is given by
equation~\eqref{eq:Xninf}. The first line in the previous expression
encodes the evolution in the absence of sources, such that for
$\Delta N < \Delta N_i$, $X_i$ keeps its value acquired during
inflation. For $\Delta N > \Delta N_i$, the source-free evolution is
decaying exponentially of exponentially fast such that $X_i$ is
completely driven by the source given in the second line of
equation~\eqref{eq:Xireh}. As a result, either $X_i$ takes its value
at the end of inflation, or the source term drives its dynamics. For
this reason, we only consider the source term from now on.

For $X_2$, plugging equation~\eqref{eq:X1reh} into \eqref{eq:Xireh},
one has
\begin{equation}
X_2(N) \simeq \dfrac{\Gammab_1}{\Hbend} e^{-\frac{2}{3} \frac{\Gammab_2}{\Hbend} e^{3 \Delta
      N /2}}  \int_{\N{end}}^N
\exp{\left(\dfrac{2}{3} \dfrac{\Gammab_2 - \Gammab_1}{\Hbend} e^{3
      \Delta y/2} + \dfrac{3}{2} \Delta y \right)} \ud y.
\label{eq:X2reh}
\end{equation}
The argument under the integral emphasizes the typical time scale $\Delta
N_{12}$ involved in the evolution of $X_2$ sourced by $X_1$. It is
defined by the transcendental equation
\begin{equation}
\dfrac{3}{2} \Delta N_{12} = \dfrac{2}{3}  \dfrac{\left |\Gammab_2 -
  \Gammab_1 \right |}{\Hbend} e^{3
      \Delta N_{12}/2},
\end{equation}
whose solution can be expressed in terms of the Lambert function
\begin{equation}
\Delta N_{12} = -\dfrac{2}{3} \Lambert{-1}\left(-\dfrac{2}{3}
\dfrac{\left|\Gammab_2 - \Gammab_1\right|}{\Hbend} \right) \simeq 
\dfrac{2}{3} \ln \left(\dfrac{3}{2} \dfrac{\Hbend}{\left|\Gammab_2 -
  \Gammab_1 \right|} \right) \simeq 
\dfrac{2}{3} \ln \left(\dfrac{\sqrt{6}}{\left|\Gammab_2 -
  \Gammab_1 \right|} \right),
\label{eq:DeltaN12}
\end{equation}
the asymptotic limit holding for $\left|\Gammab_2 - \Gammab_1\right|
\ll \Hbend$. For $1 <\Delta N < \Delta N_{12}$, we obtain the approximate
solution
\begin{equation}
X_2(1< \Delta N<\Delta N_{12}) \simeq  \dfrac{2}{3} \dfrac{\Gammab_1}{\Hbend} \exp{\left(-\dfrac{2}{3} \dfrac{\Gammab_2}{\Hbend} e^{3
      \Delta N/2} + \dfrac{3}{2} \Delta N \right)},
\label{eq:X2grow}
\end{equation}
showing that $X_2$ first exponentially grows for $\Delta N < \Delta
N_2$ and then disappears as a double decaying exponential for $\Delta
N > \Delta N_2$. However, such an evolution may be interrupted before
completion if $\Delta N_{12} < \Delta N_2$.

For $\Delta N > \Delta N_{12}$, one has to distinguish whether
$\Gammab_2 < \Gammab_1$ or $\Gammab_2 > \Gammab_1$. In the former
situation, the argument of the exponential under the integral of
equation~\eqref{eq:X2reh} becomes strongly negative and the integral
saturates for $\Delta y \simeq \Delta N_{12}$. One has
\begin{equation}
X_2 (\Delta N>\Delta N_{12}) \simeq \dfrac{2}{3} \dfrac{\Gammab_1}{\Hbend} \exp{\left(-\dfrac{2}{3} \dfrac{\Gammab_2}{\Hbend} e^{3
      \Delta N/2} + \dfrac{3}{2} \Delta N_{12} \right)}, \qquad
\Gammab_2 < \Gammab_1.
\label{eq:X2slow}
\end{equation}
This expression shows that for the ``interrupted'' case, $\Delta
N_{12} < \Delta N_2$, the exponential growth of $X_2$ is interrupted
at $\Delta N_{12}$ and $X_2$ remains constant up to $\Delta N_2$ at
which it finally disappears. From equation~\eqref{eq:DeltaN12}, the
stationary (and maximal) value of $X_2$ is
\begin{equation}
X_2 (\Delta N_{12} <\Delta N < \Delta N_2) \simeq
\dfrac{\Gammab_1}{\Gammab_1 - \Gammab_2}\,.
\label{eq:X2stat}
\end{equation}
Under the same hypothesis, $\Delta N > \Delta N_{12}$, let us discuss
the case $\Gammab_2 > \Gammab_1$. The argument of the exponential in
equation~\eqref{eq:X2reh} becomes positive and the integral blows up
as a double exponential such that one can neglect the term in $3
\Delta y/2$. The integral is now given by an exponential integral
function such that
\begin{equation}
\int_{\N{end}}^N \exp \left(\dfrac{2}{3} \dfrac{\Gammab_2 - \Gammab_1}{\Hbend}
e^{3 \Delta y /2}\right) \ud y \simeq \dfrac{2}{3} \Ei\left(\dfrac{2}{3}
\dfrac{\Gammab_2 - \Gammab_1}{\Hbend} e^{3 \Delta N/2} \right).
\end{equation}
Taking the large argument limit in the previous expression one finally
gets
\begin{equation}
X_2(\Delta N > \Delta N_{12}) \simeq \dfrac{\Gammab_1}{\Gammab_2 -
  \Gammab_1} \exp\left(-\dfrac{2}{3} \dfrac{\Gammab_1}{\Hbend} e^{3
  \Delta N/2} \right), \qquad \Gammab_2 > \Gammab_1.
\label{eq:X2fast}
\end{equation}
As one may expect, for $\Gammab_2 > \Gammab_1$, the late time
evolution of $X_2$ becomes completely driven by the evolution of $X_1$
and thus $X_2$ disappears when $\Delta N \simeq \Delta N_1$.

Equations~\eqref{eq:X2grow}, \eqref{eq:X2slow} and \eqref{eq:X2fast}
give a good approximation of the evolution of $X_2$ at all
stages. Plugging back these expressions into equation~\eqref{eq:Xireh}
would give in the same way the evolution of all $X_i$ by
recurrence. In fact, we do not need to perform any additional
calculations. Ultimately, either $\Gammab_2 < \Gammab_1$ and $X_2$
decays as in equation~\eqref{eq:X2slow}, i.e. completely disappears
for $\Delta N > \Delta N_2$; or $\Gammab_2 > \Gammab_1$ and $X_2$
decays as in equation~\eqref{eq:X2fast}, i.e. for $\Delta N > \Delta
N_1$. Both equations~\eqref{eq:X2slow} and \eqref{eq:X2fast} are of
the same functional form as equation~\eqref{eq:X1reh} such that the
ultimate behaviour of all the $X_i$ will be of the same functional
form. As a result, after some transient evolution, the component $X_i$
is expected to disappear when $\Delta N \simeq \sup\left\{\Delta N_j,
j<i \right\}$.

\begin{figure}
\begin{center}
\includegraphics[width=\wsingfig]{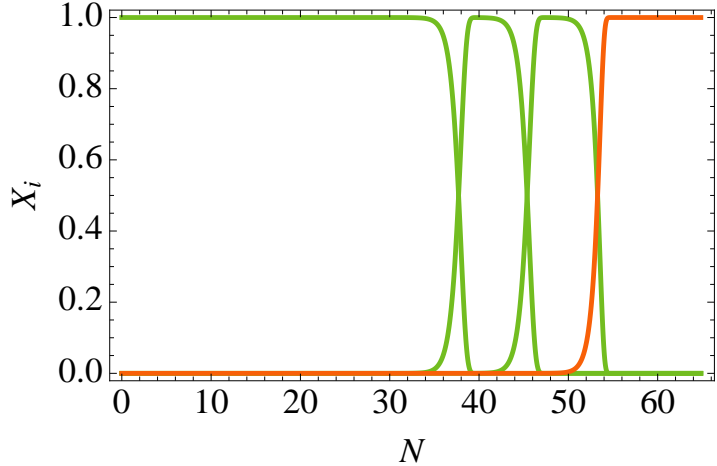}
\caption{Evolution of the relative dust fraction $X_i$ (green) and
  radiation $X$ (red) for three dust components during reheating. One
  may notice the transient stationary behaviour described by
  equation~\eqref{eq:X2stat}. Reheating ends at the time the last dust
  component of the cascade decays.}
\label{fig:X123reh}
\end{center}
\end{figure}

In figure~\ref{fig:X123reh}, we have numerically solved the evolution
equations for $n=3$ decaying dust components in addition to radiation
having $\Gamma_{i+1} < \Gamma_i$. As can be seen in this plot, the
expected behaviours of $X_i$ are recovered. While $X_1$ decays,
$X_{i>1}$ exponentially grows to reach a stationary regime and finally
disappears. Reheating ends when the Universe contains a significant
amount of radiation, namely for $X(\N{reh}) \simeq 1$. Because
\begin{equation}
X = 1 - \sum_{i=1}^n X_i,
\end{equation}
the end of reheating corresponds to the time at which the last dust
component disappears, i.e.
\begin{equation}
\Delta \N{reh} = \max\left(\Delta N_i \right) \simeq \dfrac{2}{3}
\ln \negthinspace \left[\dfrac{\sqrt{6}}{\min(\Gammab_i)}\right].
\label{eq:Nreh}
\end{equation}
Notice that only the most stable component contributes to the duration
of the reheating period, whereas, as can be seen in equation
\eqref{eq:Ninf}, the duration of inflation depends on the whole
cascade. This is the crucial feature that allows to make the model
viable by introducing the cascade. One can readily understand that
with only one dust component, both durations are linked and this was
precisely the obstruction to make the model viable with only one dust
component.

\subsubsection{BBN constraint}

\begin{figure}
\begin{center}
\includegraphics[width=1.03\wdblefig]{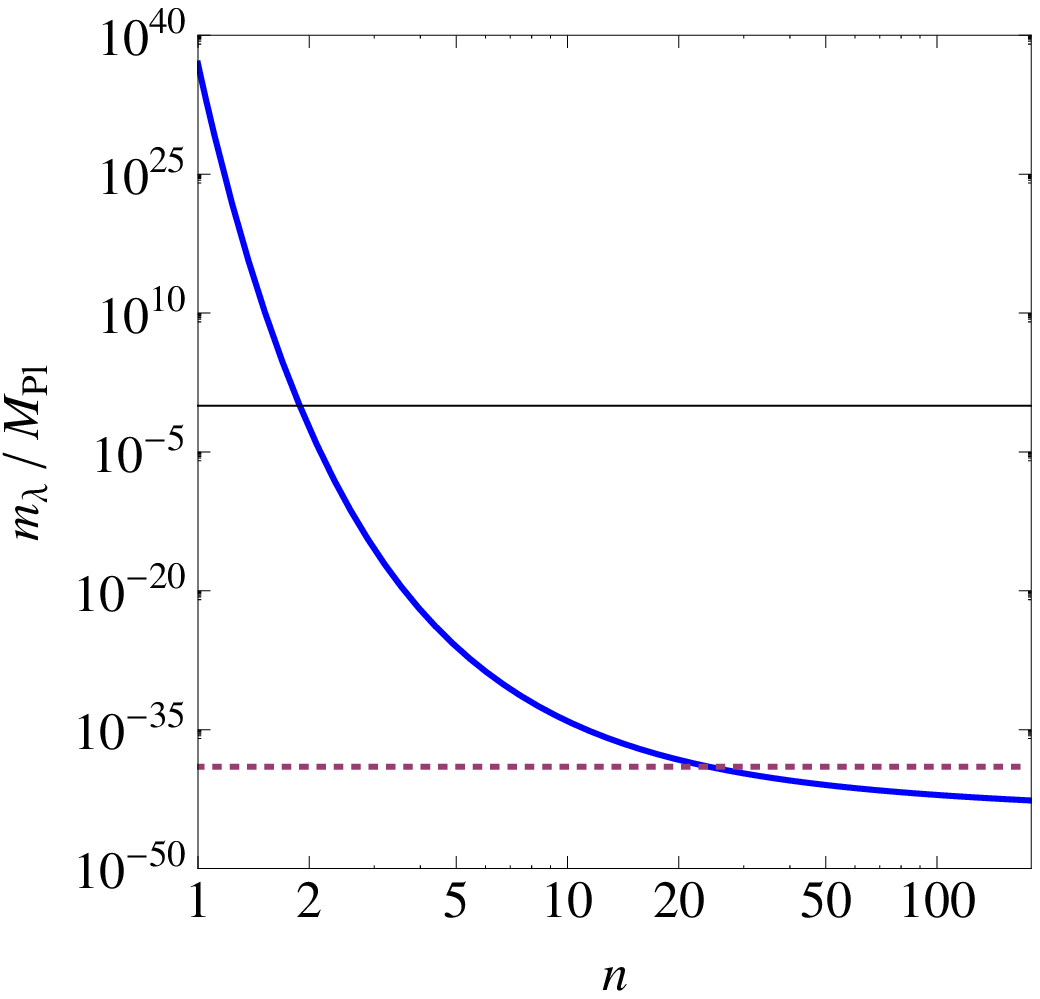}
\includegraphics[width=\wdblefig]{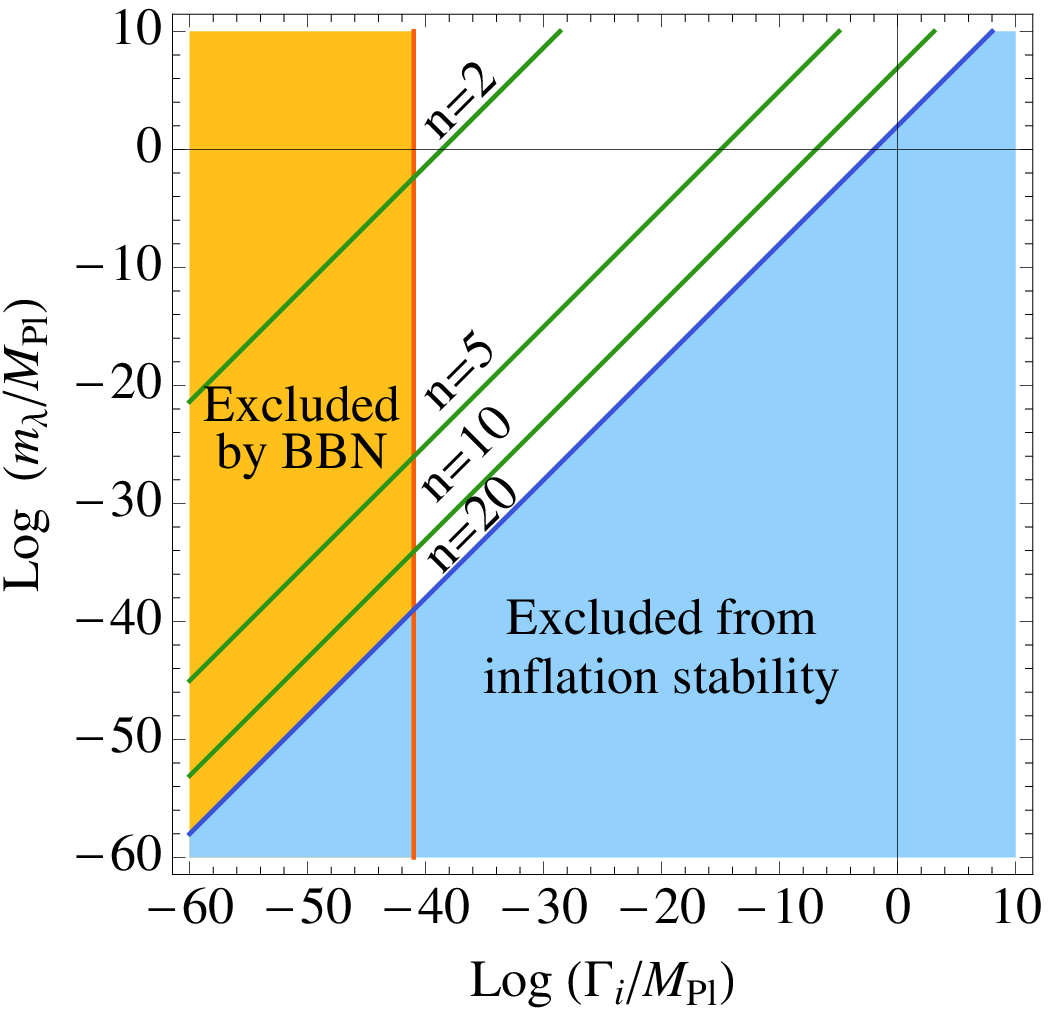}
\caption{In the left panel we show the lower bound on $\Mlambda$ from
  the constraint given in \eqref{eq:lambdabound} as a function of the
  number of dust components $n$ when they share the same decay rate
  $\Gamma_i$. The allowed region is above the curve so we see that
  only for $n=1$ the constraint can not be made below the Planck
  scale. The lower bound \eqref{eq:stability} is represented by the
  dotted line and we see that is only satisfied for $n\lesssim 20$. In
  the right panel we show the different bounds in the plane
  $(\Gamma_i,\Mlambda)$. The orange region is the bound on $\Gamma_i$
  for reheating to end before BBN. The blue region is excluded from
  the fact that the dust components need to be stable during inflation
  as expressed in \eqref{eq:stability}. Finally, the green curves
  correspond to bounds on $\Mlambda$ in order to have, at least,
  $\Delta \N{inf} \simeq 60$ e-folds of inflation.}
\label{fig:bbnbounds}
\end{center}
\end{figure}

Comparing equations~\eqref{eq:Ninf} and \eqref{eq:Nreh}, one can
obtain a long enough inflationary era to solve the usual Big-Bang
problem without having a too long reheating era. Reheating should
however end before the onset of Big-Bang Nucleosynthesis, i.e. for
$\rho_\ureh > \rho_\unuc$. According to the previous section,
\begin{equation}
\rhobreh = \rhob(\N{end}) e^{-3 \Delta \N{reh}} \simeq \dfrac{4}{3}
\left[\min(\Gammab_i) \right]^2,
\end{equation}
such that the BBN bound for $\rho_\unuc^{1/4} \simeq 10\,\MeV$ reads
\begin{equation}
\min(\Gamma_i) \ge  \dfrac{\sqrt{3 \rho_\unuc}}{2 \Mp} \simeq 10^{-41}
\,\Mp.
\end{equation}
For three fluids having the same decay rate, and $\Mlambda =
\order{\Mp}$, getting $60$ e-folds of inflation were requiring only
$\Gamma_i < 10^{-26}\,\Mlambda$ so that the BBN constraint is
trivially satisfied for that case.

In the case where all the $\Gamma_i$ are equal, one can use the BBN
bound to actually constrain the acceptable values of $\Mlambda$ in
order to have a viable inflationary model. From
equation~\eqref{eq:Ninf}, one has
\begin{equation}
3 \Delta \N{inf} \simeq - n \ln
\left(\dfrac{\Gamma_i}{\Mlambda}\right) + \dfrac{n-1}{2} \ln
\left[\dfrac{24}{(n-1)^2} \right] + \ln [(n-1)!] + \dfrac{\ln 3}{2}\,.
\end{equation}
If we saturate the minimal value for $\Gamma_i$ we can obtain the
following lower bound for $\Mlambda$ by requiring that inflation
should last longer than $\sim 60$ e-folds:
\begin{equation}
\Mlambda >  \frac{n-1}{\sqrt{24}}\left[\frac{\sqrt{8}}{(n-1)(n-1)!}\right]^{1/n}\dfrac{\sqrt{3
    \rho_\unuc}}{2 \Mp}e^{\frac{3 \Delta \N{inf}}{n}}\simeq \frac{n-1}{\sqrt{24}}\left[\frac{\sqrt{8}}{(n-1)(n-1)!}\right]^{1/n}e^{\frac{-94+3 \Delta \N{inf}}{n}}.
\label{eq:lambdabound}
\end{equation}
This bound implies that for one single dust component $n=1$ one would
need $\Mlambda\gsim 10^{36}\Mp$, well above the Planck mass. In
addition to be unnatural, we have shown earlier that the duration of
reheating would be twice longer than inflation in this case. Already
for $n=2$ we however obtain the bound $\Mlambda\gtrsim 10^{-3}\Mp$,
which is below the Planck scale. Finally, in the limit of many
intermediate species (but smaller than 180 as we required before) the
smallest possible bound is $\Mlambda \gtrsim 10^{-42}\Mp$. There is an
additional constraint that we have on $\Mlambda$ coming from the
stability of the cascade during inflation, i.e., the decay rates
should be smaller than the expansion rate during inflation for the
dust components to remain throughout the inflationary phase. This was
already stated as the condition $\Delta \N{inf}\ll\Hb/\Gammab_i$ in
order to have $X_1\simeq 1$. Thus, if we again saturate the value of
$\Gamma_i$, we obtain the additional bound
\begin{equation}
\Mlambda\gtrsim\sqrt{\frac{8}{3}} \, \Gamma_i \, \Delta\N{inf} \simeq 10^{-39}\Mp,
\label{eq:stability}
\end{equation}
where we have used $\Delta\N{inf}\simeq 60$. We see that this bound is
more stringent than the one obtained in the large $n$ limit. This
means, that there will be a maximum value for the allowed number of
species such that \eqref{eq:stability} is satisfied. Such a maximum
value is $n\lesssim 20$ as can be clearly seen in Figure
\ref{fig:bbnbounds}. In that figure we have also reported all the
bounds in the plane $(\Gamma_i,\Mlambda)$.

\section{Conclusion}
\label{sec:conclusion}

In this paper, we have discussed a scenario in which inflation is a
natural outcome of a modification of gravity in the high curvature
regime.

By considering minimal Born-Infeld gravity in the metric-affine
formalism, we have shown that accelerated expansion of the Universe
can be induced by gravitating dust at energy densities higher than
$\Mlambda^2 \Mp^2$, $\Mlambda$ being the new mass scale of the
theory. At lower curvatures, or energy-densities, GR is recovered such
that these theories automatically implement a graceful exit of the
accelerated phase without the need of scalar fields (and/or negative
pressure fluids). Graceful exit is however not enough to ensure a
viable inflationary scenario as the Universe should reheat to become
radiation dominated after inflation, and such a reheating era should
end before BBN.

Reheating within our scenario can be implemented by considering that
inflation in the Born-Infeld regime is supported by a chain of
unstable dust components that ultimately decay into radiation when the
GR regime is recovered. Because in the Palatini formalism ``the speed
of sound gravitates'' or, in other words, the background evolution not
only depends on the equation of state parameter, but also on the sound
speed, we have shown that such a scenario is severely constrained
thereby making the model relatively predictive. For instance, the
total number of e-folds cannot be made arbitrarily large, see
equation~\eqref{eq:Ninf}, and is completely fixed by the decay rate of
the dust species. As a result, observing a tiny but non-vanishing
spatial curvature today would be a natural outcome of such a scenario,
unlike in most inflationary scenarios where the typical duration is
usually much larger than $\order{10^2}$ e-folds and, therefore, any
initial spatial curvature is heavily suppressed. Another requirement
of a successful reheating phase is that at least two decaying dust
components should coexist during inflation, cascading one into
another. Indeed, when only one unstable dust specie decays into
radiation, both the duration of reheating and inflation are completely
fixed by the dust lifetime. In that situation, one obtains that
reheating lasts twice the duration of inflation, which is in tension
with cosmological observations. For more than two dust components, the
model works provided the new mass scale $\Mlambda$ is not too low (see
figure~\ref{fig:bbnbounds}), although the constraints are fairly mild
and having $\Mlambda\gsim 10^{-39}\Mp$ would suffice.

Interestingly, due to the existence of a non-vanishing speed of sound
$\cs^2$ during inflation, there is a maximum energy density
$\rhob_{\max}$, given by equation \eqref{eq:rhomax}, above which the
inflationary phase cannot exist anymore and could be
  superseded by a bounce. In addition, the presence of a positive
pressure given by the radiation component also sets an independent
upper bound on the energy density in our Born-Infeld inspired
modification of gravity. As a result, the energy densities of the
(unstable) dust components are self-regulated and cannot take
arbitrarily large values in the past. On the one hand, cascading dust
inflation provides a framework in which, at the background level,
various criticisms of the standard inflationary scenario, and
criticisms of the alternative models, seem to be addressed. On the
other hand, there are still various points to be clarified. For
instance, the nature of the dust components has not being specified
and could range from super-heavy or super-cold dark matter particles
to black holes. Being more specific on the origin of the gravitating
dust should put additional constraints on the acceptable decaying
rates, and thus on the duration of inflation and reheating. It will be
also interesting to discuss if these particles can play the role of
dark matter. Most importantly, we have not derived the primordial
power spectra that would source the cosmological perturbations. This
is indeed a non-trivial problem within the Born-Infeld class of
theories as we illustrate now.

A distinctive feature of the model is that it exhibits superinflation,
namely a negative first Hubble flow function $\epsilon_1 < 0$. Within
GR, one would immediately conclude that the spectral index of the
gravitational waves is blue, and from equation~\eqref{eq:eps1}, that
the second Hubble flow function $\epsilon_2 \simeq 3$ such that the
spectral index of the scalar modes would be super-red. However, this
is not correct as in the Born-Infeld regime the perturbations evolve
in a completely different manner than in GR. The problem for the
tensor modes $h_{ij}\equiv a^{-2} \delta g_{ij}$ has been addressed in
Ref.~\cite{Jimenez:2015caa} for a large class of Palatini theories and
it was shown that, in the absence of anisotropic stresses (as expected
in our inflationary scenario), they verify
\begin{equation}
\begin{aligned}
h_{ij} = \hT_{ij}, \qquad
\mut''_{ij} + \left(-\nabla^2 + \dfrac{\at''}{\at} \right) \mut_{ij}=0,
\end{aligned}
\label{eq:tensor}
\end{equation}
where $\hT_{ij} = \at^{-2} \delta \gt_{ij}$ stands for the tensor
perturbations of the auxiliary metric and $\mut_{ij} \equiv \at
\hT_{ij}$ is the usual quantized mode function. In the above equation
a ``prime'' denotes derivative with respect to the auxiliary conformal
time $\etat$ (defined by $\nt = \at$). The mode evolution for
$\mut_{ij}$ is identical to the one of GR, but with respect to the
auxiliary metric and coordinates. For a quasi-de Sitter expansion $\Hb
\simeq \sqrt{8/3}$ in the metric $g$, one gets $\at(\etat) \propto
\etat^{1/2}$, i.e. kination in the auxiliary metric. As a result, the
effective mass in \eqref{eq:tensor} reads $-\at''/\at = +1/(4
\eta^2)>0$ and no amplification of the tensor modes occurs. As a
result, and in spite of the superinflationary nature of the cascading
dust model, no primordial gravitational waves are generated,
independently of the energy scales involved. This is indeed a very
distinctive feature of this inflationary model. In particular,
any detection of $B$-modes in the CMB generated by primordial
gravitational waves would rule out the model.

The situation is more complex for the scalar modes and, at the time of
this writing, no conclusion can be drawn on their viability. Their
equations of motion will now involve the scalar sources, namely
perturbations in the dust components and one would need to quantize
the fluid degrees of freedom directly. This could be done either by
resorting to an effective theory for fluids~\cite{Andersson:2006nr,
  Dubovsky:2011sj, Ballesteros:2012kv, Ballesteros:2013nwa,
  Ballesteros:2014sxa} or by
computing the spectrum of thermal fluctuations. We leave however the
explicit computation for a future work.

\acknowledgments

J.B.J. acknowledges the financial support of A*MIDEX project (n¡
ANR-11-IDEX-0001-02) funded by the ``Investissements d'Avenir" French
Government program, managed by the French National Research Agency
(ANR), the Wallonia-Brussels Federation grant ARC No. 11/15-040 and
MINECO (Spain) projects FIS2011-23000 and Consolider-Ingenio MULTIDARK
CSD2009-00064.  G.J.O. is supported by a Ramon y Cajal contract, the
Spanish Grant No. FIS2011-29813-C02-02, FIS2014-57387-C3-1-P the Consolider Program
CPANPHY-1205388, and the Grant No. i-LINK0780 of the Spanish Research
Council (CSIC). This work has also been supported by CNPq (Brazilian
agency) through Project No. 301137/20 14-5.

\bibliographystyle{JHEP}
\providecommand{\href}[2]{#2}\begingroup\raggedright\endgroup


\end{document}